# Exponential Structure of Income Inequality: Evidence from 67 Countries

**Abstract:** Economic competition between humans leads to income inequality, but, so far, there has been little understanding of underlying quantitative mechanisms governing such a collective behavior. We analyze datasets of household income from 67 countries, ranging from Europe to Latin America, North America and Asia. For all of the countries, we find a surprisingly uniform rule: Income distribution for the great majority of populations (low and middle income classes) follows an exponential law. To explain this empirical observation, we propose a theoretical model within the standard framework of modern economics and show that free competition and Rawls' fairness are the underlying mechanisms producing the exponential pattern. The free parameters of the exponential distribution in our model have an explicit economic interpretation and direct relevance to policy measures intended to alleviate income inequality.

**Keywords** Income Inequality · General Equilibrium · Rawls' Fairness · Technological Progress · Entropy    **JEL Classification** D31 · D51 · D63 · E14

**Authors:**
Yong Tao[1], Xiangjun Wu[3], Tao Zhou[4], Weibo Yan[5], Yanyuxiang Huang[2], Han Yu[2], Benedict Mondal[6], Victor M. Yakovenko[7]

**Author affiliations:**
[1]College of Economics and Management, Southwest University, Chongqing 400715, China
[2]College of Hanhong, Southwest University, Chongqing 400715, China
[3]College of Economics, Hangzhou Dianzi University, Hangzhou 310018, China
[4]Big Data Research Center, University of Electronic Science and Technology of China, Chengdu 611731, China
[5]School of Economics and Finance, Xi'an Jiaotong University, Xi'an 710049, China
[6]Department of Physics, University of Maryland, College Park, MD 20742-4111, USA
[7]Department of Physics, CMTC and JQI, University of Maryland, College Park, MD 20742-4111, USA

**Acknowledgments:**
The authors would like to thank two anonymous referees and the editorial board for valuable comments and suggestions. All errors remain ours. Victor Yakovenko was supported by grant "Statistical Physics Approach to Income and Wealth Distribution" from the Institute for New Economic Thinking (INET), and Yong Tao by the Fundamental Research Funds for the Central Universities of China (Grant No. SWU1409444).

To whom correspondence should be addressed. Email: taoyingyong@yahoo.com (Yong Tao) or yakovenk@physics.umd.edu (Victor M. Yakovenko)

# 1. Introduction

Economic inequality is a universal phenomenon in human societies. Although there are broad patterns of economic inequality between countries, their sources are poorly understood and hotly debated (Kuznets 1955; Acemoglu and Robinson 2009; Autor 2014; Piketty and Saez 2014; Ravallion 2014; Nishi et al. 2015). To explain the origin of economic inequality, researchers put forward different mechanisms, such as institutional structures (Acemoglu and Robinson 2009; Piketty and Saez 2014), technological progress (Acemoglu and Robinson 2009; Autor 2014), economic growth (Ravallion 2014), psychological factor (Nishi et al. 2015), and so on. In fact, because economic inequality involves many different aspects (e.g. income, wealth, social status, and so on) in human societies, seeking a universal pattern of economic inequality seems an impossible task. Nevertheless, some researchers tried to find universal patterns in income inequality. An influential economist Vilfredo Pareto proposed that income distribution in a society is well described by a power law (Pareto 1897). Although many studies have confirmed that the high-income class of populations follows a power law (Mandelbrot 1960; Kakwani 1980), there is increasing evidence showing that it does not apply to the majority of population with lower income. Using income data for USA, Yakovenko and Rosser (2009) have shown that the US society has a well-defined two-class structure (Dragulescu and Yakovenko 2001a, 2001b; Silva and Yakovenko 2005; Yakovenko and Rosser 2009; Banerjee and Yakovenko 2010): The great majority of population (low and middle income class) follows an exponential law, while the remaining part (high income class) follows a power law. Dragulescu and Yakovenko proposed a thermal equilibrium theory based on statistical mechanics to explain the exponential pattern of income distribution (Dragulescu and Yakovenko 2000), which has won more and more support from recent empirical studies (Nirei and Souma 2007; Derzsy, Néda, and Santos 2012; Jagielski and Kutner 2013; Shaikh, Papanikolaou, and Wiener 2014; Shaikh 2016; Oancea, Andrei, and Pirjol 2016). However, it should be noted that the exponential law does not fit the super-low income data, which are usually fitted by log-normal or gamma distributions (Banerjee, Yakovenko, and Di Matteo 2006; Chakrabarti et al. 2013). Moreover, although the exponential law is quite successful in describing the low and middle income data, the mechanism of thermal equilibrium is questioned by mainstream economists (Cho 2014). These economists argue that the thermal theory of income distribution lacks solid economic foundation (Cho 2014), and so is unhelpful in making policy recommendations. In response to this criticism, we show in this paper that the exponential law of income distribution can be derived from the principles of free competition and Rawls' fairness (Rawls 1999), thus giving it a solid economic foundation (Tao 2015, 2016). Because we introduce a rigorous economic treatment, the scope of applicability of the exponential distribution is determined, and we can explain why it fails to fit super-low and high income data.

Furthermore, our results can be formulated as a powerful complement for the existing literatures. Understanding of the social impact and quantitative characterization of income inequality is a subject of great social and political importance. For the quantitative characterization of inequality, while there are plenty of case-by-case studies (Piketty and Saez 2003; Piketty 2003; Banerjee, Yakovenko, and Di Matteo 2006; Piketty and Qian 2009; Clementi, Gallegati, and Kaniadakis 2010, 2012; Jagielski and Kutner 2013; Shaikh, Papanikolaou, and Wiener 2014; Saez and Zucman 2016; Oancea, Andrei, and Pirjol 2016), most of them do not recognize the underlying universal quantitative structure of income inequality, i.e., do not see the forest for the trees. Here we present overwhelming empirical evidence, derived from the datasets for 67 countries, that the low and middle part of income distribution follows a universal exponential law. More importantly, relative to other existing distributions, the fitting parameters in our distribution have an explicit economic interpretation and direct relevance to policy measures intended to alleviate income inequality. For the social impact of inequality, there are two strands of literatures. One line focuses on how the market structure and institution influences income inequality (Katz and Autor 1999; Autor, Katz, and Kearney 2008; Heathcote, Storesletten, and Violante 2010; Moretti 2013). The other line investigates the mechanism of redistribution reducing income inequality (Piketty and Saez 2003; Piketty 2003; Atkinson, Piketty, and Saez 2011; Golosov, Maziero, and Menzio 2013; Jones 2015). In this paper, we make an attempt to combine these two lines. On one side of the empirical investigation, we show that free economy exhibits a universal two-class structure: The great majority of population (low and middle income class) follows an exponential law, while the remaining part (high income class) follows a power law. On the other side of theoretical research, we show that the exponential income structure is a result of combining free competition and Rawls' fairness, while the power income structure is due to the rule "the rich get richer" (i.e. the Matthew effect). To reduce the degree of inequality, we propose that the redistribution policy should be based on the principle of levying a tax on high-income class to pay the unemployment compensation, in line with Piketty's policy propositions.

## 2. Exponential income distribution

In fact, mathematical apparatus of modern economics has been strongly influenced by physics. Following Newton's paradigm of classical mechanics, the famous economist Leon Walras developed a set of equations that describe economic equilibrium (Walras 2003). These equations opened the paradigm of "neoclassical economics" and later were perfected by Kenneth J. Arrow and Gerard Debreu (Arrow and Debreu, 1954). Now these equations are called the "Arrow-Debreu's general equilibrium model" (ADGEM), which is the well-known standard model of modern economics (Mas-Collel, Whinston, and Green 1995). Using such a model, one can illustrate why the equilibrium allocation of social resources, in which every social member obtains maximum satisfaction, exists in an "ideal institutional environment" that ensures reasonable private property rights and judicial justice. Following a mainstream economic approach, we use ADGEM in this paper to study the equilibrium income allocation among social members. Thus, we can observe that how macro-level pattern of income inequality arises from micro-level competitive interactions of individuals embedded within an ideal institutional environment[1].

Without loss of generality, we consider an "$N$-person non-cooperative game", where are $N$ consumers (or agents), each of whom operates a firm, so there are $N$ firms. Following the basic assumptions of neoclassical economics, each consumer should be selfish and have infinite desire; therefore, all of these firms will pursue maximum profit, and all of these consumers will exchange with each other to obtain maximum satisfaction. Furthermore, if a consumer is employed in a firm that he does not operate, he will obtain the ownership share of that firm. Because consumer $i$ operates firm $i$, his income consists of the firm's operational revenue and the returns on holding the shares from other firms, where $i = 1,2, \ldots, N$. All of these settings are explained in detail in Appendix A. In accordance with the basic settings of ADGEM, all the firms should be sufficiently competitive so that monopoly cannot arise; therefore, by the rule of gaining income above, each firm actually looks like a self-employed household or a small trader. This means that we can use household income data to test validity of our upcoming model. As the Pareto optimal solution to ADGEM that captures all of these settings above, Tao proved that (Tao 2015, 2016), in the long-run competition, each consumer's equilibrium income should be completely random, and obeys the following constraint:

$$\begin{cases} I_i \geq 0 \ \ for \ \ i = 1,2, \ldots, N \\ \quad \sum_{i=1}^{N} I_i = Y \end{cases} \quad (1)$$

where $I_i$ denotes the equilibrium income of the $i$th consumer and $Y$ denotes GDP (Gross Domestic Product).

Here we use $A = \{I_1, I_2, \ldots, I_N\}$ to specify an "equilibrium income allocation" (EIA) among $N$ consumers. Due to the randomness of Pareto optimal solution (1), there is a large number of EIAs. To eliminate uncertainty of optimal allocations, the proposal of traditional economists is to seek the best one by using a social preference function (Mas-Collel, Whinston, and Green 1995). Unfortunately, Arrow's Impossibility Theorem has denied the existence of social preference (Arrow 1963). This is the well-known "dilemma of social choice". However, Tao proposes that such a dilemma can be avoided by using the paradigm of natural selection (Tao 2016). To be specific, regarding each EIA as a random event and income distribution as a set of EIAs, we make a conjecture that, among all possible income distributions, the one endowed with the largest probability, i.e. the likeliest, ought to be selected, so it is survival of the likeliest (Whitfield 2007; Harte et al. 2008; Tao 2010, 2016). If our conjecture is right, we should expect the household income data to exhibit such an income distribution.

The focus of this paper is on income distribution in a democratic economy. To find the probability of each income distribution occurring under the democratic environment, we apply Rawls' justice principle of fair equality of opportunity (Rawls 1999) to ADGEM. Since ADGEM is an ideally just procedure, fair equality of opportunity indicates that each EIA should occur with an equal probability (Tao 2016). Rawls' fairness in a democratic economy means that the door of opportunity is open to all social members (Rawls 1999). Rawls' fairness principle is illustrated for an example of "2-person allocation" in Appendix B. When Rawls' fairness principle is applied to "$N$-person allocation" subject to constraints (1) where $N$ and $Y$ are large enough, we find that the exponential income distribution occurs with the highest probability (detailed derivation is given in Appendix C):

$$\begin{cases} f(x) = \frac{1}{\theta} e^{\frac{-(x-\mu)}{\theta}} \\ x \geq \mu \end{cases} \tag{2}$$

or equivalently

$$\begin{cases} P(t \geq x) = e^{\frac{-(x-\mu)}{\theta}} \\ x \geq \mu \end{cases}. \tag{3}$$

---

[1] The emergence of income inequality can be traced back to the pioneering work of John Angle (Angle 1986, 1992, 1993, 1996, and 2006).

Here $x$ denotes income level, $f(x)$ is the probability density of income $x$, and $P(t \geq x)$ is the cumulative probability distribution, i.e. the fraction of population with the income higher than $x$.

The free parameters $\mu$ and $\theta$ denote marginal labor-capital return and marginal technology return (Tao 2010, 2016), respectively (see Appendix D). The constraint $x \geq \mu$ is considered as the Rational Agent Hypothesis (Tao 2010) in neoclassical economics, which states that firms (or agents) enter the market if and only if they can gain the marginal labor-capital return at least to pay for the cost; otherwise they will make a loss. Such a hypothesis explains why the exponential distribution fails to fit the super-low income data at $x$ lower than $\mu$. This is one limitation to applicability of the exponential income distribution (2). On the other hand, by the settings of ADGEM, each firm is sufficiently competitive, and hence looks like a self-employed household; therefore, the exponential income distribution (2) does not fit super-rich people (high income class) who should operate large firms (or monopolistic firms[2]). Thus, income distribution of super-rich people obeys a power law (Axtell 2001) due to the rule "the rich get richer" (Tao 2015) (i.e. the Matthew effect) rather than Rawls' equal opportunities. Consequently, when we fit income data using the exponential distribution (2), we should drop super-low and high income data. Finally, we point out that other scholars (Foley 1994; Chakrabarti and Chakrabarti 2009; Venkatasubramanian, Luo, and Sethuraman 2015) have also applied the concepts of Rawls' fairness, utility and maximum entropy to derive income distribution; however, our derivation has the advantage of being based on ADGEM and specifying the range of applicability of the theoretical distribution.

## 3. Empirical test for 67 countries

We can estimate the values of $\mu$ and $\theta$ by fitting empirical income data to the cumulative probability distribution given in equation (3). The datasets employed in this paper come from many sources at the country level and consist of income data for a large sample of percentiles. Using data for a wide span of years, we obtain a dataset of 67 countries around the world, especially European and Latin American countries. The sources of data are fully described in the Appendix F. Because our model is based on the ADGEM, which describes an ideal market economy, we expect the exponential distribution to be applicable for the well-developed market economies. To this end, we primarily focus on OECD countries, for which it is also easier to find detailed and reliable income distribution data. Outside OECD, it is often difficult to get detailed-enough, reliable data in the appropriate format. So, the 67 countries analyzed in this paper are those for which we managed to find the data from the sources listed in Appendix F. Further effort would be desirable to expand the list of countries in future work. From the household income data, which is classified into macro income quantile data, obtained for each country, we compute the cumulative distribution of income $P(t \geq x)$, which is the ratio of the number of social members whose income is larger than $x$ to the total population.

---
[2] In the neoclassical economics, monopolistic power implies that the behaviors among firms are highly heterogeneous. Interestingly, Lux and Marchesi (1999) also showed that heterogeneous behaviors among economic agents may lead to a power law in financial markets.

Following our theoretical construction, the empirical analysis is conducted in two steps. First, we take logs to the values of the cumulative distribution equation and run a step-by-step ordinary least square (OLS) regression to the sample. Since we investigate the relationship between cumulative distribution of income and income level, according to the scope of applicability of exponential income distribution, we need to drop the high-income samples, as they follow a power law (Axtell 2001; Tao 2015). Resorting to the goodness of fit criteria, we select the samples based on the largest adjusted $R^2$ values criteria. To be specific, we first take logs to equation (3),

$$ln[P(t \geq x)] = y = \beta x + \alpha + \varepsilon, \qquad (4)$$

where $\beta = -1/\theta$, $\alpha = \mu/\theta$, then we regress $y$ on $x$ using the OLS method. In the second step, based on the regression results obtained from the first step, we compute the value of marginal labor-capital return $\mu$, which equals to the inverse of the ratio of the intercept to the slope coefficient, that is $\mu = -\alpha/\beta$. Furthermore, by Rational Agent Hypothesis, we drop the super-low income samples whose values are less than $\mu$, and again we run an OLS regression to the "new" sample.

To illustrate our testing process, we first apply the aforementioned empirical strategy to United Kingdom. In the years of 1999-2000 to 2013-2014, following the maximal adjusted $R^2$ rule, we drop the super high income data first. According to our theoretical formulation, the high income people do not conform to the assumptions of ADGEM. In fact, the number of these people is relatively small, but their total income is quite large. When the top-income samples are removed, based on the regression parameters of equation (3), we get the value of $\mu$, then we further drop the samples whose values are less than $\mu$. Once again, we run an OLS regression on the purged data to fit the data to our exponential distribution. For comparison, we also fit the data on the full sample; see the two panels in Fig.1 for details. Likewise, the same empirical testing procedure is applied to other countries around the world. The results of fitting are shown by Figs. 2 and 3. Fig. 2 shows 34 mostly European countries for which Eurostat data are available, and Fig. 3 shows 33 countries from other areas. One can observe visually that agreement between theory and empirical data is very good. Furthermore, the goodness of fit parameters for the exponential income distribution (3) to 67 countries are reported in Tables S1-S3 (see Appendix F). We show that the adjusted $R^2$ of almost all these countries approach 0.99.

Here we point out that the method of removing the high income class using the maximized $R^2$ can be regarded as a filtering procedure. By our model, the exponential function fits the middle range of income distribution, so it is necessary to filter out the data at the high and low ends of distribution to reveal the exponential pattern. The filtering is always inevitable in any data analysis performed to extract signal from noisy or mixed data, so it is not an absolute question of data integrity, but rather a practical one of whether the filtering procedure is reasonable or not. We believe that our procedure above is reasonably reliable and convincing, because it converges after removal of a quite small fraction of the data. Later, we will observe that the estimated value of $\mu$ produced by the filtering procedure for maximized $R^2$ indeed agrees with empirical data. Despite this, we still do not verify that the estimate of $\mu$ produced by the filtering procedure is consistent. In fact, because we only collect the sample data of household income, we must prove that the estimated value of $\mu$ sufficiently approaches the true value when the number of sample is large enough; otherwise, we cannot guarantee that the estimate of $\mu$ proposed by us is consistent. In next section, we will show that the estimate of $\mu$ produced by filtering procedure is consistent.

## 4. Consistent estimate of $\mu$

In section 3, we have shown that the exponential distribution (3) remarkably fits the low and middle parts of household income data from 67 countries. The only problem is that we don't know if the fitting procedure produces the consistent estimate of $\mu$. For the full data (i.e., population), the equation (4) can be written as:
$$y_j = \beta^* x_j + \alpha^* + \varepsilon_j, \tag{5}$$
where $\beta^* = -\frac{1}{\theta^*}$, $\alpha^* = \frac{\mu^*}{\theta^*}$, and $\varepsilon_j \sim N(0, \sigma^2)$ for $j = 1, 2, \ldots, \infty$. Here $\{x_j\}_{j=1}^{\infty}$ and $\{y_j\}_{j=1}^{\infty}$ denote the full data. $\beta^*$ and $\alpha^*$ are obtained by regressing $\{y_j\}_{j=1}^{\infty}$ on $\{x_j\}_{j=1}^{\infty}$.

It must be noted that, due to the constraint $x \geq \mu$, the equation (3) differs slightly from the equation (5). Therefore, we cannot ensure if $\mu^* = \mu$. In fact, the equation (3) implies that $\{x_j\}_{j=1}^{\infty}$ should be a strictly monotonic increasing sequence with $x_j \geq 0$ for $j = 1, 2, \ldots, \infty$. More importantly, it indicates that there exists a positive integer $g^*$ to guarantee $x_k \geq \mu$ for $k = g^*, g^* + 1, \ldots, \infty$. This means that, for the full data, the equation (3) should be written as:
$$\begin{cases} y_k = \beta x_k + \alpha + \varepsilon_k \\ x_k \geq \mu \end{cases}, \tag{6}$$
where $\beta = -\frac{1}{\theta}$, $\alpha = \frac{\mu}{\theta}$, and $\varepsilon_k \sim N(0, \sigma^2)$ for $k = g^*, g^* + 1, \ldots, \infty$. Here $\beta$ and $\alpha$ are obtained by regressing $\{y_j\}_{j=g^*}^{\infty}$ on $\{x_j\}_{j=g^*}^{\infty}$.

By Lemma 4 in Appendix E, we have proved that if $g^* < \infty$, then one has $\beta = \beta^*$, $\alpha = \alpha^*$. Therefore, the equation (6) can be rewritten in the form:

$$\begin{cases} y_k = \beta^* x_k + \alpha^* + \varepsilon_k \\ x_k \geq \mu^* \end{cases}, \qquad (7)$$

where $k = g^*, g^* + 1, \ldots, \infty$ and $g^* < \infty$.

Obviously, our purpose is to find $\mu$. The equation (7) reminds us that if one can collect the full data $\{x_j\}_{j=1}^{\infty}$ and $\{y_j\}_{j=1}^{\infty}$, then $\mu$ can be obtained by computing $\mu^*$. Unfortunately, nobody can collect full data, so it's impossible to obtain the equation (7). However, based on sample data $\{x_l\}_{l=1}^n$ and $\{y_l\}_{l=1}^n$, we can consider the following statistical estimate equation:

$$\begin{cases} \hat{y}_i = \hat{\beta}_g x_i + \hat{\alpha}_g \\ x_i \geq \hat{\mu}_g \end{cases}, \qquad (8)$$

where $i = g, g+1, \ldots, n$ and $n$ denotes sample size. It's worth emphasizing that $g = g(n)$ is undetermined.

Here

$$\hat{\beta}_g = \frac{\sum_{i=g}^n (x_i - \bar{x}_g)(y_i - \bar{y}_g)}{\sum_{i=g}^n (x_i - \bar{x}_g)^2}, \qquad (9)$$

$$\hat{\alpha}_g = \bar{y}_g - \hat{\beta}_g \bar{x}_g, \qquad (10)$$

$$\hat{\mu}_g = -\frac{\hat{\alpha}_g}{\hat{\beta}_g}, \qquad (11)$$

$$\bar{x}_g = \frac{1}{n-g+1} \sum_{i=g}^n x_i, \qquad (12)$$

$$\bar{y}_g = \frac{1}{n-g+1} \sum_{i=g}^n y_i. \qquad (13)$$

Due to the absence of full data, we cannot obtain $\mu$. However, we hope $\hat{\mu}_g \to \mu$ if $n \to \infty$. In Appendix E, we have proved the following proposition:

**Proposition 3:** For a strictly monotonic increasing sequence $\{x_j\}_{j=1}^n$, if there exists an integer $g = g(n)$ to guarantee:

(A). $x_{i-1} < \mu < x_i$ or $x_i = \mu$, where $i = g < n$ and $\lim_{n \to \infty} \frac{g}{n} = 0$;

(B). $\frac{\bar{y}_g}{\hat{\beta}_g} > \delta > 0$ for any $n$;

then one has:

$$\lim_{n \to \infty} \hat{\mu}_g = \lim_{n \to \infty} \left( \bar{x}_g - \frac{\bar{y}_g}{\hat{\beta}_g} \right) = \mu, \qquad (14)$$

where $g$ is uniquely determined by $n$ and $g < \infty$. This means:

$$\lim_{n \to \infty} g = g^*. \qquad (15)$$

***Proof***. See Appendix E. □

Proposition 3 indicates that $\hat{\mu}_g$ is a consistent estimate if (A) and (B) hold. That is to say, if the sample size $n$ is large enough, we expect that $\hat{\mu}_g$ is extremely close to $\mu$. Because nobody can obtain $\mu$, our purpose can be changed to find a value close to $\mu$. Obviously, Proposition 3 implies that the estimate $\hat{\mu}_g$ will provide such a value. Next we show that (B) can be related to the correlation coefficient between $\{x_i\}_{i=g}^n$ and $\{y_i\}_{i=g}^n$.

**Lemma 5:** If $y_i < 0$ for $i = 1, \ldots, n$, and if $r_g < 0$ for any $n$, then one has $\frac{\bar{y}_g}{\hat{\beta}_g} > 0$ for any $n$, where $r_g = \frac{\sum_{i=g}^n (x_i - \bar{x}_g)(y_i - \bar{y}_g)}{\sqrt{\sum_{i=g}^n (x_i - \bar{x}_g)^2 \cdot \sum_{i=g}^n (y_i - \bar{y}_g)^2}}$ denotes the correlation coefficient between $\{x_i\}_{i=g}^n$ and $\{y_i\}_{i=g}^n$.

*Proof.* By equation (9) we have:
$$r_g = \hat{\beta}_g \cdot \sqrt{\frac{\sum_{i=g}^n (x_i - \bar{x}_g)^2}{\sum_{i=g}^n (y_i - \bar{y}_g)^2}}. \tag{16}$$

Thus, if $r_g < 0$ for any $n$, one concludes $\hat{\beta}_g < 0$ for any $n$, where we have used the Assumptions (b) and (c) in Appendix E. Since $y_i < 0$ leads to $\bar{y}_g < 0$, we conclude[3] $\frac{\bar{y}_g}{\hat{\beta}_g} > 0$ for any $n$. □

By using Lemma 5, Proposition 3 leads to the following corollary.

**Corollary 1:** For a strictly monotonic increasing sequence $\{x_j\}_{j=1}^n$, if $y_j < 0$ for $j = 1, \ldots, n$, and if there exists an integer $g = g(n)$ to guarantee:

(C). $x_{i-1} < \mu < x_i$ or $x_i = \mu$, where $i = g < n$ and $\lim_{n \to \infty} \frac{g}{n} = 0$;

(D). $r_g < \gamma < 0$ for any $n$;

then one has:
$\lim_{n \to \infty} \hat{\mu}_g = \mu$,
where $g$ is uniquely determined by $n$ and $g < \infty$. This means:
$\lim_{n \to \infty} g = g^*$.

---

[3] Here we have considered $\lim_{i \to \infty} y_i \neq 0$.

***Proof***. Using Proposition 3 and Lemma 5 we complete this proof. □

Obviously, equation (3) implies $y_i < 0$ for $i = g, g+1, \dots, n$. Therefore, we can employ the Corollary 1 to seek $\hat{\mu}_g$. The step is as below:

First, we seek the minimal $l$ to satisfy $r_l < 0$ for $\{x_i\}_{i=l}^n$ and $\{y_i\}_{i=l}^n$. Second, we regress $\{x_i\}_{i=l}^n$ and $\{y_i\}_{i=l}^n$ to obtain $h$ and $\hat{\mu}_h$. Third, we test $r_h$: If $r_h < 0$ holds, we conclude that the regress result $\hat{\mu}_h = \hat{\mu}_g$ is a valid estimate value; if $r_h \geq 0$, we use $\{x_i\}_{i=h}^n$ and $\{y_i\}_{i=h}^n$ to repeat the steps 1-3. The computing process should end at the finite steps; otherwise, $\{x_i\}_{i=1}^n$ and $\{y_i\}_{i=1}^n$ do not fit equation (3).

It's easy to check that the filtering procedure in section 3 is in accordance with the three steps above, provided that $r_g < 0$ holds. For simplicity, we only list the correlation coefficients $r_g$ for the United Kingdom in Table 1, which are all negative. The readers can test the other countries, which also exhibit the negative correlation coefficients (see Fig. 2-3). Therefore, we believe that the estimate values for $\mu$ computed in Tables S1-S3 are convincing. It's worth mentioning that the assumption $\varepsilon_j \sim N(0, \sigma^2)$ in equation (6) holds if and only if the high-income samples can be adequately removed. This is because high-income samples, which obey the power law, will lead to systematic errors so that $\varepsilon_j \sim N(0, \sigma^2)$ breaks down. In section 3, we remove the high-income samples (systematic errors) based on the rule of maximized $R^2$ to get the estimate value $\mu_R$. However, the rule of maximized $R^2$ is not the unique method. In fact, Fig. 1 implies that, for the United Kingdom, we may remove only three quantile in high-income samples to get $\hat{\mu}_g$. Remarkably, the Proposition 3 implies that $\hat{\mu}_g$ should be close to $\mu_R$ if the sample size is large enough. In terms of our data, the United Kingdom data has the most quantile, and so yields the largest sample size (approximately equals 100). Therefore, it's better to compare the estimate values $\hat{\mu}_g$ and $\mu_R$ in terms of the United Kingdom data. We have listed the results in Table 1, where the readers can check that the differences only yield the order of 0.01.

## 5. Discussion

The empirical results above imply that the exponential income law universally holds in most countries all over the world. Because we have investigated 67 countries from different areas, the validity of exponential income law appears to be robust. Compared to log-normal and gamma distributions, which have two or more fitting parameters, the exponential law essentially has only one fitting parameter $1/\theta$, and produces a more parsimonious fit of the data. More importantly, our exponential law (3) is compatible with the standard model of modern economics (namely ADGEM); therefore, the fitting parameters $\mu$ and $\theta$ have explicit economic meaning. In fact, $\mu$ denotes the marginal labor-capital return, and it is proportional to the minimum wage (Tao 2017). Concretely, we can obtain (Tao 2017):

$$\mu = \sigma \cdot \omega - \sigma \cdot r \cdot MRTS_{LK}, \tag{17}$$

where $\sigma$ denotes the marginal employment level, $\omega$ denote the minimum wage, $r$ denotes the interest rate, and $MRTS_{LK}$ denotes the marginal rate of technical substitution of labor and capital. The brief derivation for equation (17) can be found in Appendix D.

The marginal employment level $\sigma$ stands for the increasing number of employment once a firm enters markets (Tao 2017); therefore, it's easy to understand $\sigma \geq 0$. Thus, equation (17) implies that the marginal labor-capital return $\mu$ is theoretically proportional to the minimum wage $\omega$. Obviously, the minimum wage $\omega$, like unemployment compensation, can be regarded as a critical income level at which labors would like to enter or exit markets. Therefore, we might as well identify $\omega$ by the unemployment compensation.

To test the relationship between $\mu$ and $\omega$, we collected the unemployment compensation data for 26 European countries in the years of 2011 to 2014. Using the computed values of $\mu$ for European countries from Table S2, we can directly test if there is a positive relationship between marginal labor-capital return and unemployment compensation by the OLS regression. The empirical results are shown in Fig. 4 and Table S4 (see Appendix F). From these results we find that the marginal labor-capital return $\mu$ (i.e., MLCR in Fig. 4), is strongly positive correlated with the unemployment compensation (i.e., UC in Fig. 4), with the Pearson correlation coefficients being separately 0.864, 0.904, 0.899 and 0.880 (from 2011 to 2014). Remarkably, the confidence levels of correlation coefficients are surprisingly high, since $p$-value $< 0.001$ for all four years, as shown in Table S4. It is worth mentioning that the equation (17) implies that $\mu$ is inversely proportional to $r$ if[4] $MRTS_{LK} > 0$ (Tao 2017). Recently, Tao (2017) has collected the real data of the interest rate $r$ to do the cross-section regression between $\mu$, $\omega$ and $r$. Tao's empirical results show that the marginal labor-capital return $\mu$ is indeed inversely proportional to the interest rate $r$ (Tao 2017).

Due to the robust results of our study, some significant policy recommendations can be made: by moderately increasing the level of unemployment compensation, the income inequalities originated from low and middle income classes may be reduced, because the Gini coefficient of the exponential distribution is equal to $G = 1/[2(1 + \mu/\theta)]$, see detailed derivation in Tao et al (2017). To keep efficiency and fairness in competitive markets, we propose that the source of paying unemployment compensation should come from levying a tax on high income class. This is because, unlike high income class, the low and middle income class evolves to a competitive equilibrium combining efficiency and Rawls' fairness. The traditional tax policy which artificially changes the income structure of low and middle income class may harm market efficiency and fairness.

## 6. Conclusion

We have shown that the standard Arrow-Debreu's general equilibrium model combined with Rawls' fairness principle naturally produces the exponential distribution of income, which agrees well with the empirical data for 67 countries around the world. These results provide a solid justification for the exponential income distribution within the mainstream economic framework. Furthermore, our findings may have broader socio-economic implications, because the exponential income law is, effectively, a result of natural selection of the likeliest (Whitfield 2007; Tao 2016), i.e. the most probable, distribution. The Arrow-Debreu's general equilibrium model describes an ideal institutional environment (similar to ecological environment), which permits different income structures. Relative to other structures, the exponential income distribution occurs with the highest probability, and so it represents survival of the likeliest structure, also named as "Spontaneous Order" (Tao 2016). These results are relevant for evolutionary economics (Mackmurdo 1940; Nelson and Winter 1982; Potts 2001; Hodgson 2004; Dopfer 2004; Foster and Metcalfe 2012), which is concerned with the direction of social evolution. The exponential distribution (2) is obtained by maximization of entropy $ln\Omega$ (see Appendix D), which indicates the direction of evolution. According to neoclassical economics, the entropy in our model is interpreted as technological progress (Tao 2016), as discussed in Appendix D, so the higher technological progress is the likeliest direction of social evolution: among all possible social systems, those whose technological level happens to be the highest will be "selected" as survivors. In other words, those social systems that possess the lower technological level will be more likely eliminated in the process of social evolution. Our insights seem to be in accordance with the existing historical facts.

---

[4] When labor $L$ and capital $K$ substitute with each other, we have $MRTS_{LK} > 0$.

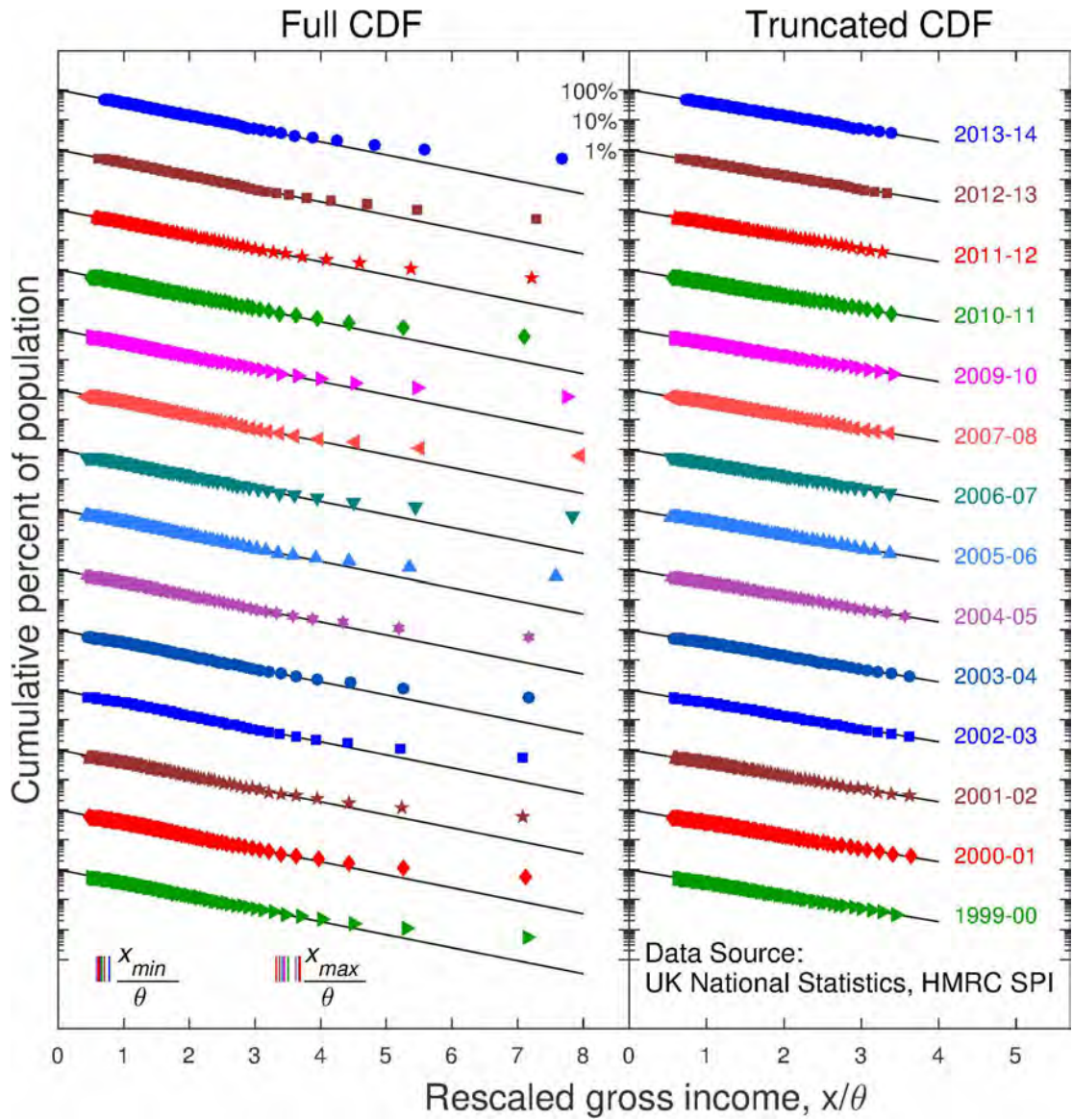

**Fig. 1. Exponential fits on full and truncated income data for the United Kingdom.** The vertical axis displays cumulative percentage of population in a logarithmic scale. The horizontal axis shows annual income rescaled by dividing by the θ-value of each year. The data and fits for each country are shifted vertically for clarity; each line of fit intersects the vertical axis at 100% population. The fitting parameters and auxiliary information regarding the fits are given in Table S1. HMRC SPI means Her Majesty's Revenue and Customs, Survey of Personal Incomes.

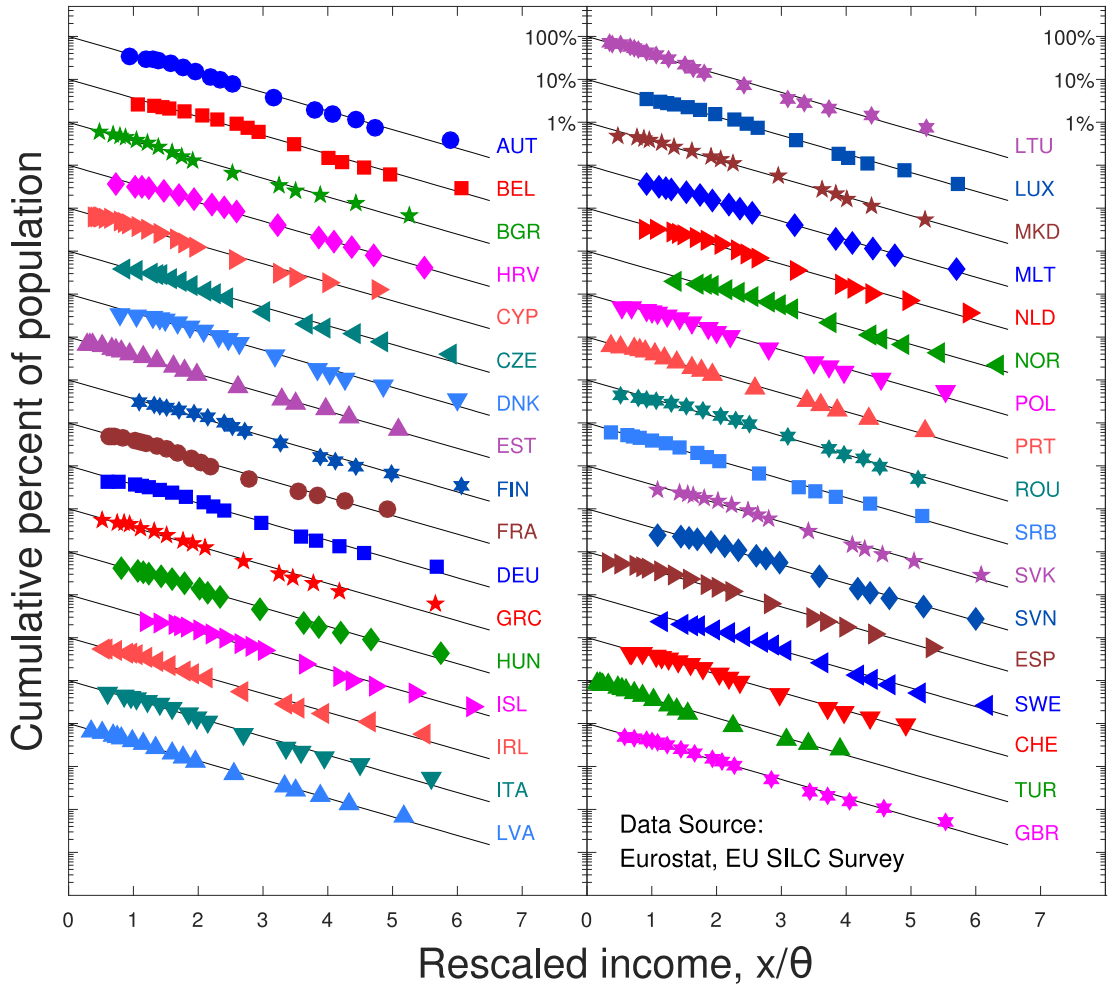

**Fig. 2. Exponential fits on truncated income data for the European Union in 2014 and its neighboring countries**. The vertical axis shows cumulative percentage of population in a logarithmic scale. The horizontal axis shows income rescaled by dividing by the corresponding θ-value of each country. The data and fits for each country are shifted vertically for clarity; each line of fit intersects the vertical axis at 100% cumulative percentage of population. The fitting parameters and auxiliary information are given in Table S2. EU-SILC means European Union Statistics on Income and Living Conditions.

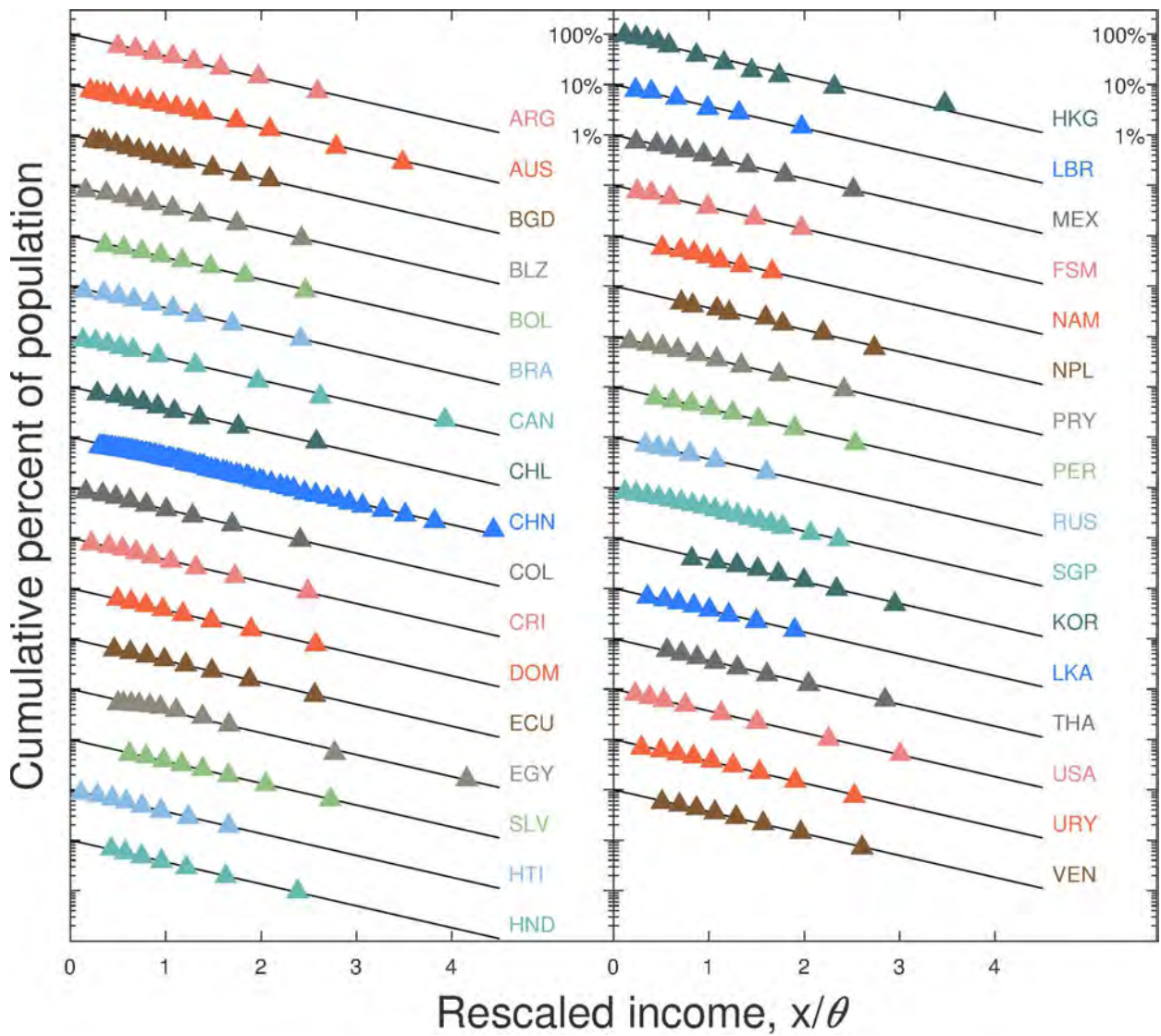

**Fig. 3. Exponential fits on truncated income distribution data for various countries over various years**. The vertical axis displays cumulative percentage of population in a logarithmic scale. The horizontal axis shows income rescaled by dividing by the corresponding θ-value of each country. The data and fits for each country are shifted vertically for clarity; each line of fit intersects the vertical axis at 100% cumulative percentage of population. The fitting parameters and auxiliary information regarding the fits are given in Table S3.

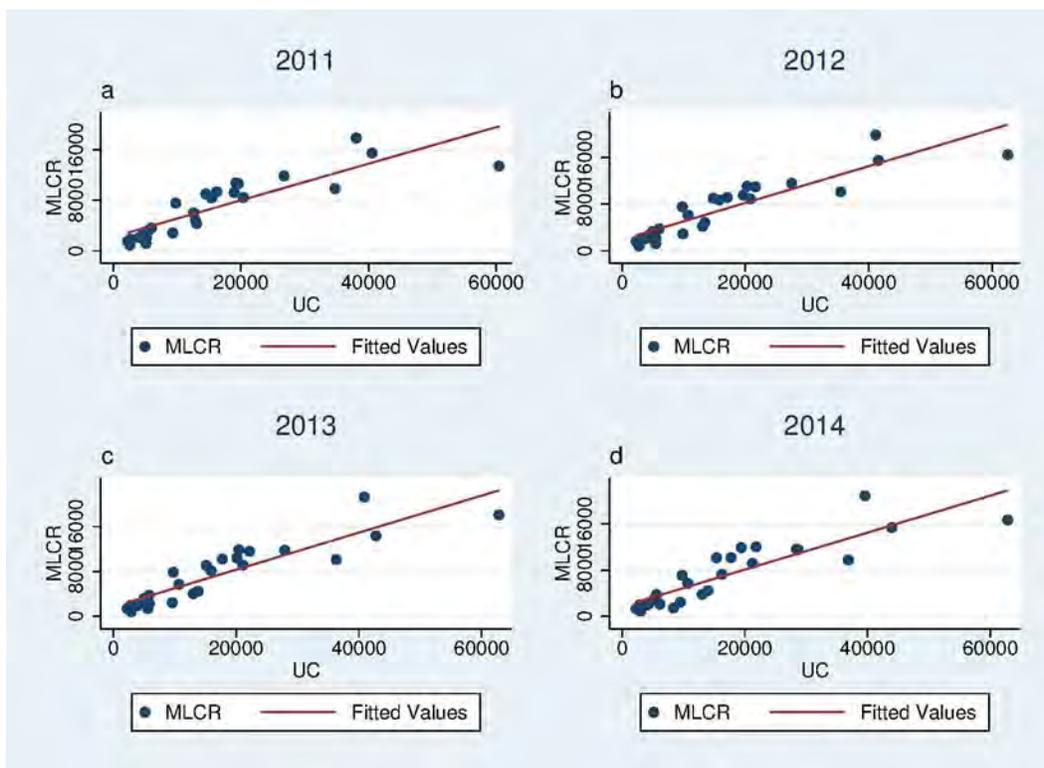

**Fig. 4. Statistical fit of the relationship between marginal labor-capital return (MLCR) and unemployment compensation (UC).** The vertical axis displays MLCR, the horizontal axis shows UC. Cross-section datasets come from 26 European countries in the year of 2011, 2012, 2013 and 2014. The local currency units (LCU) of some countries are not the euro (EUR), so annual average exchange rates, collected from Eurostat, are used to convert the UC values from LCU to EUR. The slope coefficients equal to 0.290, 0.315, 0.331, and 0.320 in the four following years, and the Pearson correlation coefficients between the two variables are separately 0.864, 0.904, 0.899 and 0.880, referring to Table S4 for details. Both the slope parameters and Pearson correlation coefficients are highly significant.

| Year | $r_g$ | $\hat{\mu}_g$ | $\mu_R$ |
|---|---|---|---|
| 1999-2000 | -0.998869 | 5662 | 5620 |
| 2000-2001 | -0.998795 | 5608 | 5538 |
| 2001-2002 | -0.998889 | 6017 | 5918 |
| 2002-2003 | -0.998872 | 6139 | 6028 |
| 2003-2004 | -0.998655 | 6167 | 6026 |
| 2004-2005 | -0.999193 | 6198 | 6090 |
| 2005-2006 | -0.999341 | 6301 | 6258 |
| 2006-2007 | -0.999275 | 6502 | 6479 |
| 2007-2008 | -0.999174 | 6790 | 6775 |
| 2009-2010 | -0.999333 | 7638 | 7644 |
| 2010-2011 | -0.999467 | 7559 | 7548 |
| 2011-2012 | -0.999281 | 8224 | 8333 |
| 2012-2013 | -0.999201 | 8869 | 9000 |
| 2013-2014 | -0.99898 | 9690 | 9906 |

**Table 1. Correlation coefficients and estimate $\hat{\mu}_g$ for the United Kingdom.** The data used reports gross annual individual income after taxes for tax years 1999-2000 to 2013-2014. $\hat{\mu}_g$ is the estimate value of $\mu$ based on removing three quantile in high-income samples, and $\mu_R$ is the estimate value of $\mu$ based on the maximized $R^2$ as reported in Table S1.

# APPENDIX

## A. *N*-person Non-cooperative Game

Arrow-Debreu's General Equilibrium Model (ADGEM) is based on the well-known two criteria of neoclassical economics: utility maximization and profit maximization. If there are $N$ consumers, each of whom operates a firm, the ADGEM describing their optimal behavior uses the following principles (Tao 2015, 2016):

(a). Profit maximization: For each firm $i = 1, \ldots, N$, $y_i^* \in Y_i$ maximizes profits such that $P \cdot y_i \leq P \cdot y_i^*$ for all $y_i \in Y_i$.

(b). Utility maximization: For each consumer $i = 1, \ldots, N$, $x_i^* \in X_i$ is the solution of maximizing the preference $\succsim_i$ under the budget set: $\{x_i \in X_i : p \cdot x_i \leq p \cdot \omega_i + \sum_{j=1}^{N} \theta_{ij} p \cdot y_j^*\}$.

(c). Market clearing: $\sum_{i=1}^{N} x_i^* = \sum_{i=1}^{N} \omega_i + \sum_{i=1}^{N} y_i^*$.

Here $x_i$ and $X_i$ represent consumption vector and consumption set of the $i$th consumer, respectively; $y_i$ and $Y_i$ represent production vector and production set of the $i$th firm, respectively (Mas-Collel, Whinston, and Green 1995); $\theta_{ij}$ represents an ownership share of each firm $j = 1, \ldots, N$ paid to the $i$th consumer. The allocation $(x_1^*, \ldots, x_N^*; y_1^*, \ldots, y_N^*)$ and a price vector $p = (p_1, \ldots, p_L)$ constitute a Pareto optimal solution to ADGEM (a)-(c).

## B. Rawls' Fairness of "2-Person Allocation"

For illustration, let us consider a simple "2-person society" in which the GDP is denoted by ＄2 and each person can earn a possible equilibrium income with ＄0, ＄1 or ＄2. For the "2-person society", the equation (1) can be expressed in the form:

$$\begin{cases} I_i = 0, 1, 2 \quad for \quad i = 1, 2 \\ \sum_{i=1}^{2} I_i = 2 \end{cases} \qquad (B.1)$$

By equation (B.1), the "2-person society" will have three equilibrium income allocation (EIA): $A_1 = \{0,2\}$, $A_2 = \{2,0\}$ and $A_3 = \{1,1\}$. They have been shown as below:

Case 1 ($A_1$)      Case 2 ($A_2$)     Case 3 ($A_3$)

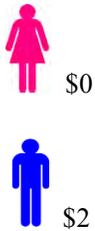 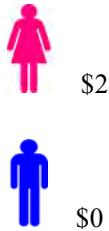 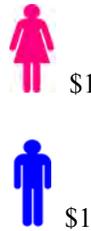

$0     $2     $1

$2     $0     $1

By Rawls' principle of fair equality of opportunity, each EIA should occur with an equal probability (Tao 2015, 2016); therefore, each person's expected income equals ＄1. The detailed calculation is as below:

Probability($A_1$)=1/3, Probability($A_2$)=1/3, Probability($A_3$)=1/3

Woman's Expected Income = 0×(1/3) + 2×(1/3) + 1×(1/3)=1      (B.2)

Man's Expected Income = 2×(1/3) + 0×(1/3) + 1×(1/3)=1      (B.3)

This means that each person owns the equal opportunity of earning money. If we denote the equal income distribution by $a$ and the unequal income distribution by $b$, we do have $a = \{A_3\}$ and $b = \{A_1, A_2\}$. By Rawls' principle of fair equality of opportunity, $a$ will occur with probability $1/3$ and $b$ will occur with probability $2/3$. Following the rule of "survival of the likeliest", $b$ will be a result of natural selection.

## C. Density Function of Income Distribution

For "$N$-person allocation", Tao has shown that, by applying Rawls' fairness into equation (1) where $N$ and $Y$ are large enough, one will get the exponential income distribution which occurs with the highest probability (Tao 2015, 2016):

$$a_k = g_k e^{-\frac{(\varepsilon_k - \mu)}{\theta}}, \tag{C.1}$$

$\varepsilon_1 < \varepsilon_2 < \cdots < \varepsilon_n$.

Here $\mu$ and $\theta$ denote marginal labor-capital return and marginal technology return, respectively (Tao 2016); readers can find the origin of these two parameters in Appendix D.

The formula (C.1) indicates that there are $a_k$ consumers each of which obtains $\varepsilon_k$ units of revenue, and $k$ runs from 1 to $n$. Because income distribution (C.1) will occur with the highest probability, Tao call it the "spontaneous economic order" (Tao 2016).

The formula (C.1) can be rewritten in the form of continuous function. To see this, let us first observe:

$$\sum_{k=1}^{n} a_k = N, \tag{C.2}$$

which leads to:

$$\sum_{k=1}^{n} \frac{a_k}{N} = 1. \tag{C.3}$$

Here $\frac{a_k}{N}$ denotes the proportion of populations each of whom earns $\varepsilon_k$ units of income.

Now we write $\frac{a_k}{N}$ in the form of continuous function: $f(x)$. To this end, let us order:

$$f(x) = w \cdot e^{-\frac{(x-\mu)}{\theta}}, \tag{C.4}$$

where $x$, which replaces $\varepsilon_k$, denotes a continuous income level, and by Rational Agent Hypothesis one has (Tao 2010) $x \geq \mu$.

Here $w$ is an undetermined constant, which will be determined by the sum formula (C.3).

Let us replace $\frac{a_k}{N}$ by (C.4), and transform sum operation of formula (C.3) into integral operation:

$$\int_{\mu}^{+\infty} w \cdot e^{\frac{-(x-\mu)}{\theta}} dx = 1, \tag{C.5}$$

which leads to $w = \frac{1}{\theta}$.

Finally, we obtain the density function of income distribution:

$$f(x) = \frac{1}{\theta} e^{\frac{-(x-\mu)}{\theta}}. \tag{C.6}$$

## D. Technological Progress and Entropy

Because the firm consists of labor and capital, the Cobb-Douglas aggregate production function (or GDP) of neoclassical economics can be written in the form (Tao 2010, 2016):

$$Y = Y(N(L,K), H), \tag{D.1}$$

where $L$ and $K$ denote labor and capital, whereas $N$ and $H$ denote the number of firms and technological progress.

The complete differential of (D.1) yields [see also Eq. (9) in Banerjee and Yakovenko (2010)]:

$$dY(N(L,K), H) = \mu dN(L,K) + \theta dH, \tag{D.2}$$

where $\mu = \partial Y/\partial N$ and $\theta = \partial Y/\partial H$ denote the marginal labor-capital return and the marginal technology return (Tao 2016), respectively.

Here Tao identifies the entropy $ln\Omega$ with the technological progress $H$ (Tao 2010, 2016):

$$H = ln\Omega, \tag{D.3}$$

where $\Omega$ denotes the number of equilibrium income allocations that a given income distribution contains (Furthermore, $\Omega$ also measures the choice freedom of social members (Tao 2016)). For example, for the 2-person society described by Appendix B, we have $\Omega(a) = 1$ and $\Omega(b) = 2$. By maximizing (Tao 2010, 2015, 2016) $\Omega$ one can obtain the exponential income distribution (2). Consequently, the technological progress $H$ can be regarded as the entropy of socio-economical systems.

Furthermore, the complete differential of equation (D.1) can be rewritten in the form:

$$dY = \omega dL + rdK + \theta dH, \tag{D.4}$$

where $\omega = \partial Y/\partial L$ and $r = \partial Y/\partial K$ denote marginal labor return and marginal capital return (Tao 2017), respectively. On the one hand, we might as well assume that capital markets exhibit perfect competition, so $r$ also denotes the interest rate. On the other hand, by the principle of diminishing marginal return in neoclassical economics, $\omega$ denotes the minimum wage. Comparing equations (D.2) and (D.4), we can obtain (Tao 2017):

$$\mu = \omega \cdot \sigma - r \cdot \sigma \cdot MRTS_{LK}, \tag{D.5}$$

where $\sigma = dL/dN$ and $MRTS_{LK} = -dK/dL$. Here $\sigma$ denotes the marginal employment level and $MRTS_{LK}$ denotes the marginal rate of technical substitution of labor and capital (Tao 2017).

## E. Main Propositions

To obtain the consistent estimate of $\mu$, we do the estimate analysis in terms of two cases: full sample and truncation sample. In this paper, $lim_{n\to\infty} a_n = a$ means $lim_{n\to\infty} P(a_n = a) = 1$, where $P(\xi)$ denotes the probability of $\xi$ occurring.

### E1. Full Sample

Let us first drop the constraint $x \geq \mu$. For the full data (i.e., population), the equations (4) can be written in the form:
$$y_j = \beta^* x_j + \alpha^* + \varepsilon_j, \tag{E.1}$$
$$\mu^* = -\frac{\alpha^*}{\beta^*}, \tag{E.2}$$
where $\beta^* = -\frac{1}{\theta^*}$, $\alpha^* = \frac{\mu^*}{\theta^*}$, and $\varepsilon_j \sim N(0, \sigma^2)$ for $j = 1, 2, \ldots, \infty$. Here $\{x_j\}_{j=1}^{\infty}$ and $\{y_j\}_{j=1}^{\infty}$ denote the full data. $\beta^*$ and $\alpha^*$ are obtained by regressing $\{y_j\}_{j=1}^{\infty}$ on $\{x_j\}_{j=1}^{\infty}$.

For the full sample[5], the sample estimates of equations (E.1) and (E.2) yield:
$$\hat{y}_i = \hat{\beta} x_i + \hat{\alpha}, \tag{E.3}$$
$$\hat{\mu} = -\frac{\hat{\alpha}}{\hat{\beta}}, \tag{E.4}$$
where $i = 1, 2, \ldots, n$.

Due to the absence of the constraint $x \geq \mu$, equation (E.1) differs slightly from equation (3); therefore, we don't ensure if $\mu^* = \mu$. In section E2, we will discuss the estimate of $\mu$ when $x \geq \mu$ holds. In this section, we mainly investigate the consistency of the estimate (E.4).

Taking the least squares estimation on equation (E.3) we have:
$$\hat{\beta} = \frac{\sum_{i=1}^{n}(x_i - \bar{x})(y_i - \bar{y})}{\sum_{i=1}^{n}(x_i - \bar{x})^2}, \tag{E.5}$$
$$\hat{\alpha} = \bar{y} - \hat{\beta}\bar{x}, \tag{E.6}$$
where $\bar{x} = \frac{1}{n}\sum_{i=1}^{n} x_i$ and $\bar{y} = \frac{1}{n}\sum_{i=1}^{n} y_i$.

Since the exponential distribution (3) is only suitable for the low and middle parts of the income data, we should drop the high income data. Moreover, due to the economic meanings of $x_i$ in the equation (3), $\{x_i\}_{i=1}^{n}$ should be a monotonic increasing sequence. Thus, we can make the following assumptions.

**Assumptions:** (a). $|x_i| < \infty$ and $|y_i| < \infty$ for $i = 1, 2, \ldots, n$.
               (b). $\{x_i\}_{i=1}^{n}$ is strictly monotonic increasing sequence with $x_i \geq 0$ for $i = 1, 2, \ldots, n$.

---

[5] Full sample means $\{x_1, \ldots, x_n\}$, where $n$ denotes the sample size.

(c). $\varepsilon_j$ are i.i.d. $N(0, \sigma^2)$.

Next we verify that $\hat{\beta}$ and $\hat{\alpha}$ are consistent estimates.

**Theorem 1:** Assume that $\varepsilon_j$ are i.i.d. $N(0, \sigma^2)$. If there is $lim_{n\to\infty}(X^TX)^{-1} = \mathbf{0}$, then one has:

$$lim_{n\to\infty} \hat{\beta} = \beta^*, \tag{E.7}$$
$$lim_{n\to\infty} \hat{\alpha} = \alpha^*, \tag{E.8}$$

where $X = \begin{pmatrix} x_1 & \cdots & x_n \\ 1 & \cdots & 1 \end{pmatrix}^T$.

***Proof.*** See Lai, Robbins, and Wei (1979). □

To verify equations (E.7) and (E.8), we can only prove the following proposition.

**Proposition 1:** $lim_{n\to\infty}(X^TX)^{-1} = \mathbf{0}.$

***Proof.*** It's easy to compute:

$$(X^TX)^{-1} = \frac{1}{n\sum_{i=1}^n x_i^2 - (\sum_{i=1}^n x_i)^2} \begin{pmatrix} n & -\sum_{i=1}^n x_i \\ -\sum_{i=1}^n x_i & \sum_{i=1}^n x_i^2 \end{pmatrix},$$

so proving $lim_{n\to\infty}(X^TX)^{-1} = \mathbf{0}$ is equivalent to verifying:

$$lim_{n\to\infty} \frac{1}{n\sum_{i=1}^n x_i^2 - (\sum_{i=1}^n x_i)^2} \begin{pmatrix} n & -\sum_{i=1}^n x_i \\ -\sum_{i=1}^n x_i & \sum_{i=1}^n x_i^2 \end{pmatrix} = \begin{pmatrix} 0 & 0 \\ 0 & 0 \end{pmatrix}. \tag{E.9}$$

Obviously, proving equation (E.9) is equivalent to verifying the following three equations:

$$lim_{n\to\infty} \frac{n}{n\sum_{i=1}^n x_i^2 - (\sum_{i=1}^n x_i)^2} = 0, \tag{E.10}$$

$$lim_{n\to\infty} \frac{\sum_{i=1}^n x_i}{n\sum_{i=1}^n x_i^2 - (\sum_{i=1}^n x_i)^2} = 0, \tag{E.11}$$

$$lim_{n\to\infty} \frac{\sum_{i=1}^n x_i^2}{n\sum_{i=1}^n x_i^2 - (\sum_{i=1}^n x_i)^2} = 0. \tag{E.12}$$

One can compute:

$$n\sum_{i=1}^n x_i^2 - (\sum_{i=1}^n x_i)^2 = n^2 \left[\frac{1}{n}\sum_{i=1}^n x_i^2 - (\bar{x})^2\right]. \tag{E.13}$$

Furthermore, we have the following result:

$$\frac{1}{n}\sum_{i=1}^n x_i^2 - (\bar{x})^2 = \frac{1}{n}\sum_{i=1}^n x_i^2 - 2(\bar{x})^2 + (\bar{x})^2 = \frac{1}{n}\sum_{i=1}^n (x_i - \bar{x})^2. \tag{E.14}$$

By Assumption (b) we must have $\sum_{i=1}^n (x_i - \bar{x})^2 \neq 0$; otherwise, $x_i = \bar{x}$ for $i = 1, 2, \ldots, n$, contradicting the strict monotonicity. On the other hand, by the strict monotonicity, there should be at most one number $x_l$ leading to $x_l = \bar{x}$. Thus, if we order $\min_{i \neq l}|x_i - \bar{x}| = A$, then we have $\sum_{i=1}^{n}(x_i - \bar{x})^2 \geq 0 + (n-1) \cdot A^2$.

Consequently, by equation (E.14) we can obtain:

$$\left|\frac{1}{n}\sum_{i=1}^{n} x_i^2 - (\bar{x})^2\right| = \left|\frac{1}{n}\sum_{i=1}^{n}(x_i - \bar{x})^2\right| \geq \frac{n-1}{n} \cdot A^2. \tag{E.15}$$

Using equations (E.13) and (E.15) one has

$$\left|\frac{1}{n\sum_{i=1}^{n} x_i^2 - (\sum_{i=1}^{n} x_i)^2}\right| = \left|\frac{1}{n^2\left[\frac{1}{n}\sum_{i=1}^{n} x_i^2 - (\bar{x})^2\right]}\right| \leq \frac{1}{n^2 \cdot \frac{n-1}{n} A^2} = \frac{1}{n \cdot (n-1) \cdot A^2}. \tag{E.16}$$

On the other hand, by Assumption (a), we can order $max_i|x_i| = B$; therefore, we have:
$$\left|\sum_{i=1}^{n} x_i\right| = \sum_{i=1}^{n}|x_i| \leq n \cdot B, \tag{E.17}$$

$$\left|\sum_{i=1}^{n} x_i^2\right| = \sum_{i=1}^{n} x_i^2 \leq n \cdot B^2. \tag{E.18}$$

Using equations (E.16)-(E.18), we can obtain:

$$\left|\frac{n}{n\sum_{i=1}^{n} x_i^2 - (\sum_{i=1}^{n} x_i)^2}\right| \leq \frac{n}{n \cdot (n-1) \cdot A^2} = \frac{1}{(n-1) \cdot A^2}. \tag{E.19}$$

$$\left|\frac{\sum_{i=1}^{n} x_i}{n\sum_{i=1}^{n} x_i^2 - (\sum_{i=1}^{n} x_i)^2}\right| \leq \frac{n \cdot B}{n \cdot (n-1) \cdot A^2} = \frac{B}{(n-1) \cdot A^2}. \tag{E.20}$$

$$\left|\frac{\sum_{i=1}^{n} x_i^2}{n\sum_{i=1}^{n} x_i^2 - (\sum_{i=1}^{n} x_i)^2}\right| \leq \frac{n \cdot B^2}{n \cdot (n-1) \cdot A^2} = \frac{B^2}{(n-1) \cdot A^2}. \tag{E.21}$$

Imposing $n \to \infty$ on equations (E.19)-(E.21) one can obtain equations (E.10)-(E.12). □

By using the Theorem 1, it's easy to compute:
$$lim_{n \to \infty} \hat{\mu} = -\frac{\alpha^*}{\beta^*} = \mu^*. \tag{E.22}$$

Equation (E.22) indicates that if there is no the constraint $x \geq \mu$, then the estimate $\hat{\mu}$ is consistent. However, the existence of the constraint $x \geq \mu$ may lead to the inconsistency of estimate $\hat{\mu}$.

### E2. Truncation sample

Now let us recover the constraint $x \geq \mu$. Since the constraint $x \geq \mu$ holds, we attempt to construct a truncation estimate of $\mu$. To this end, we might as well assume that $\mu$ has existed. Thus, the truncation of the full data $x_j$ can be written as:
$$x_j \geq \mu, \tag{E.23}$$
where $j = g^*, g^* + 1, \ldots, \infty$.

Using the truncation data (E.23), equation (4) can be written as:
$$y_k = \beta x_k + \alpha + \varepsilon_k, \tag{E.24}$$
$$x_k \geq \mu, \tag{E.25}$$
where $\beta = -\frac{1}{\theta}$, $\alpha = \frac{\mu}{\theta}$, and $\varepsilon_k \sim N(0, \sigma^2)$ for $k = g^*, g^* + 1, \ldots, \infty$. Here $\beta$ and $\alpha$ are

obtained by regressing $\{y_j\}_{j=g^*}^{\infty}$ on $\{x_j\}_{j=g^*}^{\infty}$.

Thus, the sample estimates of equations (E.24) and (E.25) yield:

$$\hat{y}_i = \hat{\beta}_g x_i + \hat{\alpha}_g, \tag{E.26}$$

$$x_i \geq \hat{\mu}_g, \tag{E.27}$$

where $i = g, g+1, \ldots, n$ and $g = g(n)$. Here $\{x_i\}_{i=g}^n$ and $\{y_i\}_{i=g}^n$ denote truncation sample.

It's worth emphasizing that $g^*$ and $g = g(n)$ are undetermined.

Taking the least squares estimation on equation (E.26) we have:

$$\hat{\beta}_g = \frac{\sum_{i=g}^n (x_i - \bar{x}_g)(y_i - \bar{y}_g)}{\sum_{i=g}^n (x_i - \bar{x}_g)^2}, \tag{E.28}$$

$$\hat{\alpha}_g = \bar{y}_g - \hat{\beta}_g \bar{x}_g, \tag{E.29}$$

where $\bar{x}_g = \frac{1}{n-g+1} \sum_{i=g}^n x_i$ and $\bar{y}_g = \frac{1}{n-g+1} \sum_{i=g}^n y_i$.

The main purpose of this section is to derive the estimate $\hat{\mu}_g$. Assume $g^* < \infty$, thus we will have the following theorem and proposition:

**Theorem 2:** Assume that $\varepsilon_j$ are i.i.d. $N(0, \sigma^2)$. If there is $\lim_{n\to\infty} (X_{g^*}^T X_{g^*})^{-1} = \mathbf{0}$, then one has:

$$\lim_{n\to\infty} \hat{\beta}_{g^*} = \beta, \tag{E.30}$$

$$\lim_{n\to\infty} \hat{\alpha}_{g^*} = \alpha, \tag{E.31}$$

where $X_{g^*} = \begin{pmatrix} x_{g^*} & \cdots & x_n \\ 1 & \cdots & 1 \end{pmatrix}^T$.

*Proof.* Same as the Theorem 1. □

**Proposition 2:** $\lim_{n\to\infty} (X_{g^*}^T X_{g^*})^{-1} = \mathbf{0}.$

*Proof.* Same as the Proposition 1. □

Consistent with the form of equation (E.4), $\hat{\mu}_g$ can be defined as:

$$\hat{\mu}_g = -\frac{\hat{\alpha}_g}{\hat{\beta}_g}. \tag{E.32}$$

Now we start to derive the consistent condition of guaranteeing the validity of estimate (E.32).

Substituting equation (E.29) into (E.32) one has:

$$\hat{\mu}_g = \bar{x}_g - \frac{\bar{y}_g}{\hat{\beta}_g}, \tag{E.33}$$

which guarantees that the constraint of equation (E.26) has been imposed on the estimate (E.32).

On the other hand, equation (E.27) indicates:
$$\bar{x}_g > \hat{\mu}_g + \delta, \tag{E.34}$$
where we have used the Assumption (b) and $\delta > 0$.

Inserting equation (E.33) into equation (E.34) yield:
$$\frac{\bar{y}_g}{\hat{\beta}_g} > \delta > 0, \tag{E.35}$$

which guarantees that the constraint of equation (E.27) has been imposed on the estimate (E.32).

Thus, we can obtain the core proposition of this Appendix as below:

**Proposition 3:** For a strictly monotonic increasing sequence $\{x_j\}_{j=1}^n$, if there exists an integer $g = g(n)$ to guarantee:

(A). $x_{i-1} < \mu < x_i$ or $x_i = \mu$, where $i = g < n$ and $lim_{n \to \infty} \frac{g}{n} = 0$;

(B). $\frac{\bar{y}_g}{\hat{\beta}_g} > \delta > 0$ for any $n$;

then one has:
$$lim_{n \to \infty} \hat{\mu}_g = lim_{n \to \infty} \left( \bar{x}_g - \frac{\bar{y}_g}{\hat{\beta}_g} \right) = \mu, \tag{E.36}$$
where $g$ is uniquely determined by $n$ and $g < \infty$. This means:
$$lim_{n \to \infty} g = g^*. \tag{E.37}$$

To verify the Proposition 3, we need to prove the following four lemmas:

**Lemma 1:** If $\{\xi_i\}_{i=1}^n$ is a monotonic sequence and if $|\xi_i| < \infty$ for any $i$, then one has:
$$lim_{n \to \infty} \xi_n = \xi, \tag{E.38}$$
where $|\xi| < \infty$.

*Proof.* See the theorem 3.14 in Rudin (1976). □

**Lemma 2:** For the sequence $\{\xi_i\}_{i=1}^n$, if $lim_{n \to \infty} \xi_n = \xi$, then one has:
$$lim_{n \to \infty} \frac{1}{n} \sum_{i=1}^n \xi_i = \xi. \tag{E.39}$$

*Proof.* Since $lim_{n \to \infty} \xi_n = \xi$, by the definition of limit, for every $\epsilon > 0$ there always exists a positive integer $N$ so that when $k > N$, one has:
$$|\xi_k - \xi| < \frac{\epsilon}{2}. \tag{E.40}$$

To verify equation (E.39), we only need to prove:

$$lim_{n\to\infty}\left(\frac{1}{n}\sum_{i=1}^{n}\xi_i - \xi\right) = 0; \tag{E.41}$$

that is, for every $\epsilon > 0$ there always exists a positive integer $N_0$ so that when $n > N_0$, one has:

$$\left|\frac{1}{n}\sum_{i=1}^{n}\xi_i - \xi\right| < \epsilon. \tag{E.42}$$

It's easy to compute:

$$\left|\frac{1}{n}\sum_{i=1}^{n}\xi_i - \xi\right|$$

$$= \left|\frac{1}{n}\left[\sum_{i=1}^{N}(\xi_i - \xi) + \sum_{j=N+1}^{n}(\xi_j - \xi)\right]\right|$$

$$\leq \frac{1}{n}\left|\sum_{i=1}^{N}(\xi_i - \xi)\right| + \frac{1}{n}\left|\sum_{j=N+1}^{n}(\xi_j - \xi)\right|. \tag{E.43}$$

Because $lim_{n\to\infty}\xi_n = \xi$, it's easy to verify that $|\xi_i| < \infty$ and $|\xi| < \infty$. Thus, one has $max_i|\xi_i - \xi| < \infty$. Consequently, thanks to $j > N$, equation (E.43) can be written in the form:

$$\left|\frac{1}{n}\sum_{i=1}^{n}\xi_i - \xi\right|$$

$$\leq \frac{N}{n}max_i|\xi_i - \xi| + \frac{n-N}{n}\frac{\epsilon}{2}$$

$$< \frac{N}{n}max_i|\xi_i - \xi| + \frac{\epsilon}{2}. \tag{E.44}$$

where we have used equation (E.40).

It's easy to compute $lim_{n\to\infty}\frac{N}{n}max_i|\xi_i - \xi| = 0$. This means that for every $\epsilon > 0$ there always exists a positive integer $N_1$ so that when $k > N_1$, one has:

$$\frac{N}{k}max_i|\xi_i - \xi| < \frac{\epsilon}{2}. \tag{E.45}$$

Let us order $N_0 = max\{N, N_1\}$; thus, substituting equation (E.45) into equation (E.44) we conclude that for every $\epsilon > 0$ when $n > N_0$, there always holds:

$$\left|\frac{1}{n}\sum_{i=1}^{n}\xi_i - \xi\right| < \epsilon. \quad \square$$

**Lemma 3:** If $lim_{n\to\infty}\frac{g}{n} = 0$, one has:

$$lim_{n\to\infty}\bar{x}_g = lim_{n\to\infty}\bar{x} = x,$$
$$lim_{n\to\infty}\bar{y}_g = lim_{n\to\infty}\bar{y} = y,$$

where $x = lim_{n\to\infty}x_n$ and $y = \beta^*x + \alpha^*$.

***Proof.*** We first verify $lim_{n\to\infty}\bar{x}_g = lim_{n\to\infty}\bar{x}$. It's easy to check:

$$\bar{x} = \frac{1}{n}\sum_{i=1}^{n}x_i = \frac{1}{n}\sum_{i=1}^{g-1}x_i + \frac{n-g+1}{n}\frac{1}{n-g+1}\sum_{j=g}^{n}x_j = \frac{1}{n}\sum_{i=1}^{g-1}x_i + \frac{n-g+1}{n}\bar{x}_g$$

(E.46)

Since $lim_{n\to\infty} \frac{g}{n} = 0$, imposing $n \to \infty$ on equation (E.46) one obtains:

$lim_{n\to\infty} \bar{x}_g = lim_{n\to\infty} \bar{x}$,

where we have used $|x_i| < \infty$.

Since Assumptions (a) and (b) hold, by using Lemma 1 one has: $lim_{n\to\infty} x_n = x$. This means that by using Lemma 2 one obtains $lim_{n\to\infty} \bar{x} = x$. Therefore, we verify $lim_{n\to\infty} \bar{x}_g = lim_{n\to\infty} \bar{x} = x$.

Now we start to verify $lim_{n\to\infty} \bar{y}_g = lim_{n\to\infty} \bar{y} = y$.

Based on the same technique from equation (E.46), by Assumption (a) we can verify $lim_{n\to\infty} \bar{y}_g = lim_{n\to\infty} \bar{y}$. By using equation (E.1), one has:

$$\bar{y} = \beta^* \bar{x} + \alpha^* + \bar{\varepsilon}, \tag{E.47}$$

where $\bar{\varepsilon} = \frac{1}{n}\sum_{i=1}^{n} \varepsilon_i$.

By Assumption (c) $\varepsilon_j$ are i.i.d. $N(0, \sigma^2)$, so by using the law of large numbers, it's easy to obtain:

$$lim_{n\to\infty} \bar{\varepsilon} = E(\varepsilon_i | x_1, \dots, x_n) = 0. \tag{E.48}$$

Therefore, substituting equation (E.48) into equation (E.47) leads to:

$lim_{n\to\infty} \bar{y} = \beta^* lim_{n\to\infty} \bar{x} + \alpha^* + lim_{n\to\infty} \bar{\varepsilon} = \beta^* x + \alpha^*$. □

**Lemma 4:** If $lim_{n\to\infty} \frac{g}{n} = 0$, one has:

$$lim_{n\to\infty} \hat{\beta}_g = lim_{n\to\infty} \hat{\beta} = \beta = \beta^*. \tag{E.48}$$

$$lim_{n\to\infty} \hat{\alpha}_g = lim_{n\to\infty} \hat{\alpha} = \alpha = \alpha^*. \tag{E.49}$$

*Proof.* Here we only verify equation (E.48). By the same technique, one can verify equation (E.49).

It's easy to check:

$$\hat{\beta} = \frac{\sum_{i=1}^{n}(x_i-\bar{x})(y_i-\bar{y})}{\sum_{i=1}^{n}(x_i-\bar{x})^2} = \frac{\frac{1}{n}\sum_{i=1}^{g-1}(x_i-\bar{x})(y_i-\bar{y}) + \frac{1}{n}\sum_{j=g}^{n}(x_j-\bar{x})(y_j-\bar{y})}{\frac{1}{n}\sum_{i=1}^{g-1}(x_i-\bar{x})^2 + \frac{1}{n}\sum_{j=g}^{n}(x_j-\bar{x})^2}. \tag{E.50}$$

Imposing $n \to \infty$ on equation (E.50) one obtains:

$$lim_{n\to\infty} \hat{\beta} = \frac{lim_{n\to\infty}\frac{1}{n}\sum_{j=g}^{n}(x_j-\bar{x})(y_j-\bar{y})}{lim_{n\to\infty}\frac{1}{n}\sum_{j=g}^{n}(x_j-\bar{x})^2}.$$

$$= \frac{lim_{n\to\infty}\frac{1}{n}\sum_{j=g}^{n}(x_j-lim_{n\to\infty}\bar{x})(y_j-lim_{n\to\infty}\bar{y})}{lim_{n\to\infty}\frac{1}{n}\sum_{j=g}^{n}(x_j-lim_{n\to\infty}\bar{x})^2}, \tag{E.51}$$

where we have used $|x_i| < \infty$, $|y_i| < \infty$ and $lim_{n\to\infty} \frac{g}{n} = 0$.

Using Lemma 3, equation (E.51) can be rewritten as:

$$lim_{n\to\infty} \hat{\beta} = \frac{lim_{n\to\infty} \frac{1}{n}\sum_{j=g}^{n}(x_j - lim_{n\to\infty} \bar{x}_g)(y_j - lim_{n\to\infty} \bar{y}_g)}{lim_{n\to\infty} \frac{1}{n}\sum_{j=g}^{n}(x_j - lim_{n\to\infty} \bar{x}_g)^2}.$$

$$= lim_{n\to\infty} \frac{\sum_{j=g}^{n}(x_j - \bar{x}_g)(y_j - \bar{y}_g)}{\sum_{j=g}^{n}(x_j - \bar{x}_g)^2}$$

$$= lim_{n\to\infty} \hat{\beta}_g. \tag{E.52}$$

Because $g^* < \infty$, by the same technique for deriving equation (E.52), we can obtain: $lim_{n\to\infty} \hat{\beta} = lim_{n\to\infty} \hat{\beta}_{g^*}$.

On the other hand, by Theorem 1 one has $lim_{n\to\infty} \hat{\beta} = \beta^*$ and by Theorem 2 one has $lim_{n\to\infty} \hat{\beta}_{g^*} = \beta$. Therefore, we conclude that equation (E.48) holds. □

Now we start to verify the Proposition 3.

***Proof of Proposition 3.*** Imposing $n \to \infty$ on equation (E.33) one obtains:

$$lim_{n\to\infty} \hat{\mu}_g = lim_{n\to\infty} \bar{x}_g - \frac{lim_{n\to\infty} \bar{y}_g}{lim_{n\to\infty} \hat{\beta}_g}. \tag{E.53}$$

Using Lemmas 1-4, equation (E.53) equals:

$$lim_{n\to\infty} \hat{\mu}_g = x - \frac{y}{\beta}. \tag{E.54}$$

We have known

$$\mu = -\frac{\alpha}{\beta}. \tag{E.55}$$

Imposing $n \to \infty$ on equation (E.29) one obtains:
$$\alpha = y - \beta x. \tag{E.56}$$
Substituting equations (E.55) and (E.56) into equation (E.54) yields:

$$lim_{n\to\infty} \hat{\mu}_g = x - \frac{y}{\beta} = \mu. \tag{E.57}$$

Because $\frac{\bar{y}_g}{\hat{\beta}_g} > \delta > 0$ for any $n$, by Lemmas 3 and 4 we have:

$$lim_{n\to\infty} \frac{\bar{y}_g}{\hat{\beta}_g} = \frac{y}{\beta} \geq \delta > 0. \tag{E.58}$$

Thus, by equation (E.57) we must conclude:
$$\mu < x < \infty, \tag{E.59}$$
where we have used $|x_i| < \infty$.

Since $x_{g-1} < \mu < x_g$ or $x_g = \mu$, by Assumption (b) we have:

$0 \leq \mu < x < \infty$.

Now we further verify that, for a given $n$, there is no another $g' \neq g$ to guarantee that $x_{g'-1} < \mu < x_{g'}$ or $x_{g'} = \mu$. We discuss this point in terms of two cases. First, if $x_{g'-1} < \mu < x_{g'}$ holds, we have to conclude $x_{g-1} < \mu < x_{g'}$ and $x_{g'-1} < \mu < x_g$. For this case, we might as well assume $g' > g$, which by Assumption (b) leads to $x_{g'} > x_g$. This means $x_g \leq x_{g-1}$, contradicting Assumption (b). Likewise, we can refute $g' < g$. Second, if $x_{g'} = \mu$ and $g' \neq g$, then by Assumption (b) the contradiction occurs. In summary, we must conclude $g' = g$.

Finally, we verify $g < \infty$. If $g = \infty$, by $x_{g-1} < \mu < x_g$ or $x_g = \mu$, we have to conclude $lim_{l \to \infty} x_l = \mu = x$, which contradicts $\mu < x < \infty$.

Based on the results above, we should have $lim_{n \to \infty} g = g^*$. To see this, we might as well assume that $lim_{n \to \infty} g > g^*$. Then, by $x_{g-1} < \mu = x_{g^*} < x_g$, one has $lim_{n \to \infty} g - 1 = g^*$, which contradicts $x_{lim_{n \to \infty} g - 1} < x_{g^*}$, where we have used $g < \infty$. □

# F. Description of Data Sources

| Source | Countries | Link |
|---|---|---|
| Socio-Economic Database of Latin America and the Caribbean | ARG, BLZ, BOL, BRA, CHL, COL, CRI, DOM, ECU, SLV, HTI, HND, MEX, PRY, PER, URY, VEN | http://sedlac.econo.unlp.edu.ar/eng/statistics.php |
| Australian Bureau of Statistics | AUS | http://www.ausstats.abs.gov.au/ausstats/subscriber.nsf/0/B0530ECF7A48B909CA257BC80016E4D3/$File/65230_2011-12.pdf |
| Eurostat | AUT, BEL, BGR, HRV, CYP, CZE, DNK, EST, FIN, FRA, DEU, GRC, HUN, ISL, IRL, ITA, LVA, LTU, LUX, MKD, MLT, NLD, NOR, POL, PRT, ROU, SRB, SVK, SVN, ESP, SWE, CHE, TUR, GBR | http://appsso.eurostat.ec.europa.eu/nui/show.do?dataset=ilc_di01&lang=en |
| Statistics Canada | CAN | http://www.statcan.gc.ca/tables-tableaux/sum-som/l01/cst01/famil105a-eng.htm |
| Hong Kong | HKG | http://www.census2011.gov.hk/pdf/household-income.pdf |
| Nepal Rastra Bank | NPL | http://www.nrb.org.np/red/publications/study_reports/Study_Reports--Household%20Budget%20Survey%202008%20(Report).pdf |
| Russian Federal State Statistics Service | RUS | http://www.arcticstat.org/Table.aspx/Region/Russian_Federation/Indicator/Personal!Household_Income/2008-08-21-05/10874 |
| Singapore Department of Statistics | SGP | http://www.singstat.gov.sg/docs/default-source/default-document-library/publications/publications_and_papers/household_income_and_expenditure/pp-s22.pdf |
| Korean Statistical Information Service | KOR | http://kosis.kr/statHtml/statHtml.do?orgId=101&tblId=DT_1L6E001&conn_path=I2&language=en |

| National Statistical Office of Thailand | THA | http://web.nso.go.th/en/survey/house_seco/data/Whole%20Kingdom_13_FullReport.pdf |
| --- | --- | --- |
| United Kingdom National Statistics | GBR | https://www.gov.uk/government/uploads/system/uploads/attachment_data/file/503472/SPI_National_Statistics_T3_1_to_T3_11.pdf |
| United States Census Bureau | USA | http://www2.census.gov/programs-surveys/demo/tables/p60/252/table3.pdf |
| Eurostat | BGR, CZE, DNK, HUN, ISL, LTU, LVA, NOR, POL, SWE | http://ec.europa.eu/eurostat/web/exchange-rates/data/database |
| OECD Statistics | AUT, BEL, BGR, CZE, CHE, DEU, DNK, ESP, EST, FIN, FRA, GRC, HUN, IRL, ISL, ITA, LTU, LUX, LVA, NLD, NOR, POL, PRT, SVK, SVN, SWE | http://stats.oecd.org/Index.aspx?DataSetCode=FIXINCLSA# |
| Federated States of Micronesia | FSM | http://prism.spc.int/images/documents/HEIS/2005_FSM_HIES_Report-Final.pdf |
| Department of Census and Statistics Ministry of Finance and Planning Sri Lanka | LKA | http://www.statistics.gov.lk/HIES/HIES2012PreliminaryReport.pdf |
| Bangladesh Bureau of Statistics - Ministry of Planning | BGD | http://catalog.ihsn.org/index.php/catalog/2257 |
| Liberia Institute for Statistics and Geo-Information Services - Government of Liberia | LBR | http://microdata.worldbank.org/index.php/catalog/2563 |
| Central Agency for Public Mobilization and Statistics | EGY | http://www.ilo.org/surveydata/index.php/catalog/1261 |
| (CAPMAS) - Arab Republic of Egypt | | |
| Namibia Statistics Agency | NAM | http://www.ilo.org/surveydata/index.php/catalog/320 |
| China Institute for Income Distribution | CHN | http://www.ciidbnu.org/ |

| Years | θ | μ | $x_{min}$ | $x_{max}$ | % < $x_{min}$ | % > $x_{max}$ | $R^2_{adj}$ |
|---|---|---|---|---|---|---|---|
| 1999-00 | 9,436 | 5,620 | 10,100 | 47,300 | 6 | 6 | 0.9988 |
| 2000-01 | 9,965 | 5,538 | 9,060 | 46,000 | 6 | 5 | 0.9985 |
| 2001-02 | 10,488 | 5,918 | 8,360 | 45,300 | 6 | 5 | 0.9988 |
| 2002-03 | 10,614 | 6,028 | 7,580 | 45,800 | 6 | 5 | 0.9987 |
| 2003-04 | 10,749 | 6,026 | 7,690 | 46,000 | 7 | 5 | 0.9985 |
| 2004-05 | 11,351 | 6,090 | 7,000 | 43,600 | 6 | 5 | 0.9992 |
| 2005-06 | 12,031 | 6,258 | 6,710 | 41,800 | 6 | 6 | 0.9996 |
| 2006-07 | 12,432 | 6,479 | 6,470 | 40,400 | 6 | 6 | 0.9995 |
| 2007-08 | 12,958 | 6,775 | 6,190 | 40,500 | 6 | 6 | 0.9995 |
| 2009-10 | 13,542 | 7,644 | 6,190 | 39,000 | 5 | 6 | 0.9995 |
| 2010-11 | 13,548 | 7,548 | 6,070 | 38,400 | 5 | 6 | 0.9996 |
| 2011-12 | 13,846 | 8,333 | 5,970 | 38,000 | 4 | 7 | 0.9997 |
| 2012-13 | 13,861 | 9,000 | 5,610 | 36,200 | 4 | 7 | 0.9997 |
| 2013-14 | 13,930 | 9,906 | 5,740 | 32,600 | 3 | 7 | 0.9997 |

**Table S1**. **Fitting parameters and auxiliary information for the data in Fig. 1**. θ and μ are given in pounds sterling (GBP). % below $x_{min}$ and % above $x_{max}$, respectively, show the percentage of the population below $x_{min}$ and above $x_{max}$. $R^2_{adj}$ are the coefficients of determination for the fits. The data used reports gross annual individual income after taxes for tax years 1999-2000 to 2013-2014. As this is taxpayer data, income values below a minimum threshold are not reported in the survey.

| Country | Code | θ | μ | $x_{min}$ | $x_{max}$ | % < $x_{min}$ | % > $x_{max}$ | $R_{adj}^2$ | LCU/EUR |
|---|---|---|---|---|---|---|---|---|---|
| Austria | AUT | 13,173 | 10,147 | 12,393 | 77,527 | 10 | 1 | 0.992 | 1 |
| Belgium | BEL | 10,570 | 10,178 | 11,354 | 64,128 | 10 | 1 | 0.987 | 1 |
| Bulgaria | BGR | 2,728 | 882 | 1,276 | 14,346 | 10 | 1 | 0.993 | 1.95702 |
| Croatia | HRV | 3,082 | 2,002 | 2,278 | 16,898 | 10 | 1 | 0.995 | 7.575604 |
| Cyprus | CYP | 11,829 | 5,477 | 5,628 | 56,541 | 3 | 2 | 0.988 | 1 |
| Czech Rep | CZE | 4,403 | 3,677 | 3,777 | 25,948 | 5 | 1 | 0.988 | 25.98018 |
| Denmark | DNK | 14,858 | 11,620 | 11,781 | 89,144 | 5 | 1 | 0.985 | 7.456595 |
| Estonia | EST | 6,309 | 2,093 | 2,131 | 32,128 | 4 | 1 | 0.998 | 1 |
| Finland | FIN | 12,295 | 11,569 | 13,368 | 74,520 | 10 | 1 | 0.989 | 1 |
| France | FRA | 13,950 | 9,190 | 9,278 | 68,570 | 4 | 2 | 0.989 | 1 |
| Germany | DEU | 12,425 | 7,288 | 7,580 | 70,646 | 5 | 1 | 0.990 | 1 |
| Greece | GRC | 5,771 | 2,163 | 3,000 | 32,633 | 10 | 1 | 0.980 | 1 |
| Hungary | HUN | 2,830 | 1,893 | 2,315 | 16,261 | 10 | 1 | 0.991 | 296.8946 |
| Iceland | ISL | 10,140 | 12,042 | 12,111 | 63,103 | 5 | 1 | 0.990 | 162.378 |
| Ireland | IRL | 14,353 | 7,093 | 7,765 | 78,783 | 4 | 1 | 0.990 | 1 |
| Italy | ITA | 11,223 | 4,464 | 6,864 | 62,775 | 10 | 1 | 0.987 | 1 |
| Latvia | LVA | 4,500 | 1,443 | 1,557 | 23,297 | 5 | 1 | 0.997 | 1 |
| Lithuania | LTU | 4,477 | 1,311 | 1,520 | 23,476 | 4 | 1 | 0.991 | 3.45189 |
| Luxembou | LUX | 19,627 | 15,385 | 18,167 | 112,232 | 10 | 1 | 0.995 | 1 |
| Macedonia | MKD | 1,450 | 558 | 689 | 7,569 | 10 | 1 | 0.996 | 76 |
| Malta | MLT | 7,364 | 5,996 | 6,728 | 41,930 | 10 | 1 | 0.997 | 1 |
| Netherland | NLD | 11,232 | 9,771 | 10,051 | 66,190 | 5 | 1 | 0.993 | 1 |
| Norway | NOR | 18,668 | 20,908 | 25,527 | 118,483 | 10 | 1 | 0.991 | 7.806466 |
| Poland | POL | 3,728 | 2,001 | 2,038 | 20,633 | 5 | 1 | 0.993 | 4.197272 |
| Portugal | PRT | 6,823 | 2,428 | 2,514 | 35,553 | 5 | 1 | 0.996 | 1 |
| Romania | ROU | 1,423 | 682 | 742 | 7,276 | 10 | 1 | 0.989 | 4.429688 |
| Serbia | SRB | 1,969 | 536 | 712 | 10,206 | 10 | 1 | 0.996 | 112.963 |
| Slovakia | SVK | 3,381 | 3,010 | 3,685 | 20,566 | 10 | 1 | 0.992 | 1 |
| Slovenia | SVN | 5,612 | 5,697 | 6,157 | 33,721 | 10 | 1 | 0.992 | 1 |
| Spain | ESP | 9,862 | 3,806 | 5,098 | 52,593 | 10 | 1 | 0.996 | 1 |
| Sweden | SWE | 12,341 | 11,909 | 14,217 | 76,208 | 10 | 1 | 0.988 | 8.651489 |
| Switzerlan | CHE | 24,014 | 16,718 | 20,518 | 118,006 | 10 | 2 | 0.995 | 1.231088 |
| Turkey | TUR | 8,732 | 1,282 | 1,400 | 34,122 | 3 | 2 | 0.990 | 2.337995 |
| United Kin | GBR | 14,286 | 7,942 | 8,166 | 79,048 | 5 | 1 | 0.993 | 0.806064 |

**Table S2**. **Fitting parameters for the data in Fig. 2**. ISO 3166-1 alpha-3 codes are used as the country codes. The θ and μ-values are given in EUR. $x_{min}$ and $x_{max}$, respectively, show the minimum and maximum income values for each country in EUR used in the statistical fit. % below $x_{min}$ and % above $x_{max}$, respectively, show the percentage of the population sample below $x_{min}$ and above $x_{max}$. $R_{adj}^2$ are the coefficients of determination for the fits. The last column contains exchange rates from LCU to EUR. The data used reports disposable annual household income equivalized according to the modified-OECD scale. All countries' data are for 2014, with the exception of Turkey, for which 2012 data was the most current available.

| Country | Code | Year | θ | μ | $x_{min}$ | $x_{max}$ | % < $x_{min}$ | % > $x_{max}$ | $R_{adj}^2$ | LCU/USD |
|---|---|---|---|---|---|---|---|---|---|---|
| Argentina | ARG | 2013 | 3,687 | 1,056 | 836 | 4348 | 20 | 10 | 0.9960 | 5.46 |
| Australia | AUS | 2012 | 74,244 | 14,201 | 299 | 4999 | 3 | 4 | 0.9946 | 0.97 |
| Bangladesh | BGD | 2010 | 1,457 | 111 | 2000 | 17500 | 5 | 16 | 0.9977 | 69 |
| Belize | BLZ | 1999 | 928 | 135 | 24 | 371 | 10 | 10 | 0.9963 | 1.98 |
| Bolivia | BOL | 2012 | 1,344 | 209 | 282 | 1921 | 20 | 10 | 0.9949 | 6.96 |
| Brazil | BRA | 2012 | 3,601 | 440 | 85 | 1457 | 10 | 10 | 0.9962 | 2.01 |
| Canada | CAN | 2012 | 37,546 | 2,793 | 5000 | 150000 | 7 | 2 | 0.9965 | 1.02 |
| Chile | CHL | 2011 | 3,823 | 818 | 42308 | 385052 | 10 | 10 | 0.9941 | 468.50 |
| China | CHN | 2013 | 6,957 | 1,967 | 13000 | 189596 | 8 | 2 | 0.9966 | 6.14 |
| Colombia | COL | 2012 | 2,245 | 217 | 53848 | 801705 | 10 | 10 | 0.9990 | 1780.00 |
| Costa Rica | CRI | 2012 | 3,917 | 599 | 34957 | 400274 | 10 | 10 | 0.9975 | 492.55 |
| Dominican Re | DOM | 2011 | 1,538 | 384 | 2396 | 12518 | 20 | 10 | 0.9999 | 37.95 |
| Ecuador | ECU | 2012 | 1,517 | 344 | 58 | 324 | 20 | 10 | 0.9991 | 1.00 |
| Egypt | EGY | 2012 | 3,303 | 1,404 | 9000 | 75000 | 3 | 3 | 0.9852 | 6.04 |
| El Salvador | SLV | 2012 | 1,039 | 407 | 53 | 236 | 20 | 10 | 0.9983 | 1.00 |
| Haiti | HTI | 2001 | 179 | 0 | 38 | 581 | 10 | 20 | 0.9972 | 23.50 |
| Honduras | HND | 2011 | 1,124 | 46 | 757 | 4220 | 30 | 10 | 0.9983 | 18.90 |
| Hong Kong | HKG | 2011 | 26,635 | 623 | 2000 | 59999 | 2 | 4 | 0.9736 | 7.78 |
| Liberia | LBR | 2014 | 2,029 | 285 | 3500 | 30000 | 9 | 17 | 0.9900 | 90 |
| Mexico | MEX | 2012 | 1,753 | 377 | 461 | 4906 | 10 | 10 | 0.9966 | 13.35 |
| Micronesia | FSM | 2005 | 10,146 | 403 | 2500 | 19999 | 6 | 17 | 0.9983 | 1 |
| Namibia | NAM | 2009 | 4,004 | 1,279 | 15701 | 51311 | 10 | 30 | 0.9915 | 7.708 |
| Nepal | NPL | 2006 | 475 | 228 | 2064 | 7954 | 20 | 10 | 0.9854 | 73.50 |
| Paraguay | PRY | 2011 | 2,138 | 293 | 119143 | 1700749 | 10 | 10 | 0.9973 | 3950.00 |
| Peru | PER | 2012 | 1,910 | 448 | 184 | 1075 | 20 | 10 | 0.9954 | 2.66 |
| Russia | RUS | 2006 | 3,348 | 857 | 2500 | 12000 | 8 | 27 | 0.9973 | 26.86 |
| Singapore | SGP | 2014 | 81,540 | 3,783 | 1000 | 19999 | 12 | 10 | 0.9955 | 1.25 |
| South Korea | KOR | 2014 | 15,419 | 9,045 | 1065687 | 3834176 | 20 | 10 | 0.9845 | 1011.74 |
| Sri Lanka | LKA | 2012 | 2,756 | 688 | 10750 | 57499 | 10 | 20 | 0.9972 | 131.9 |
| Thailand | THA | 2013 | 1,732 | 400 | 2481 | 12608 | 25 | 8 | 0.9977 | 30.67 |
| United State | USA | 2014 | 66,526 | 5,693 | 15000 | 199999 | 13 | 6 | 0.9997 | 1.00 |
| Uruguay | URY | 2012 | 4,628 | 1,343 | 2297 | 19903 | 10 | 10 | 0.9936 | 20.45 |
| Venezuela | VEN | 2006 | 1,177 | 327 | 106889 | 549483 | 20 | 10 | 0.9975 | 2150.00 |

**Table S3. Fitting parameters for the data in Fig. 3**. ISO 3166-1 alpha-3 codes are used as the three-letter country abbreviations. θ and μ are given in United States dollars (USD). $x_{min}$ and $x_{max}$, respectively, show the minimum and maximum income values for each country in USD used in the statistical fit. % below $x_{min}$ and % above $x_{max}$, respectively, show the percentage of the population below $x_{min}$ and above $x_{max}$. $R_{adj}^2$ are the coefficients of determination for the fits. Exchange rates from LCU to USD are given in the last column. As the data are obtained from various sources, there are differences in the methodologies of data collection. Some data sources report income values that are equivalized, while others report gross household income, while others still do not specify. Some sources report annual income, others monthly, and one weekly. All numerical values are therefore rescaled to show annual values.

**Table S4a**

| Country | Code | 2011 μ | uc | adj uc | LCU/EUR | 2012 μ | uc | adj uc | LCU/EUR | 2013 μ | uc | ad uc | LCU/EUR | 2014 μ | uc | adj uc | LCU/EUR |
|---|---|---|---|---|---|---|---|---|---|---|---|---|---|---|---|---|---|
| Austria | AUT | 9011.558 | 14507 | 14507 | 1 | 8996.557 | 14791 | 14791 | 1 | 9089.328 | 15136 | 15136 | 1 | 10147 | 15400 | 15400 | 1 |
| Belgium | BEL | 9376.624 | 16205 | 16205 | 1 | 9161.577 | 17069 | 17069 | 1 | 10234.03 | 17759 | 17759 | 1 | 10178 | 17759 | 17759 | 1 |
| Estonia | EST | 2048.152 | 5184 | 5184 | 1 | 2121.868 | 5502 | 5502 | 1 | 2274.793 | 5866 | 5866 | 1 | 2093.1 | 6169 | 6169 | 1 |
| Finland | FIN | 10778.19 | 19204 | 19204 | 1 | 10981.69 | 21636 | 21636 | 1 | 11599.92 | 22230 | 22151 | 1 | 11569 | 28694 | 28694 | 1 |
| France | FRA | 8456.686 | 20375 | 20375 | 1 | 8884.214 | 20821 | 20821 | 1 | 9157.157 | 21147 | 21128 | 1 | 9189.8 | 21265 | 21265 | 1 |
| Germany | DEU | 8391.587 | 15418 | 15418 | 1 | 8721.566 | 15745 | 15745 | 1 | 8156.623 | 15927 | 15927 | 1 | 7288.1 | 16272 | 16272 | 1 |
| Greece | GRC | 3247.839 | 5538 | 5538 | 1 | 2638.16 | 4320 | 4320 | 1 | 2257.006 | 4320 | 4320 | 1 | 2162.9 | 4320 | 4320 | 1 |
| Ireland | IRL | 7575.375 | 9776 | 9776 | 1 | 7527.011 | 9776 | 9776 | 1 | 7884 | 9776 | 9776 | 1 | 7093.2 | 9776 | 9776 | 1 |
| Italy | ITA | 4805.254 | 12879 | 12879 | 1 | 4848.9 | 13432 | 13432 | 1 | 4466.387 | 13835 | 13835 | 1 | 4463.8 | 13987 | 13987 | 1 |
| Luxembourg | LUX | 15500.1 | 40539 | 40539 | 1 | 15508.71 | 41577 | 41577 | 1 | 14383.48 | 42904 | 42788 | 1 | 15385 | 43936 | 43936 | 1 |
| Netherlands | NLD | 9865.709 | 34715 | 34715 | 1 | 10137.4 | 35447 | 35447 | 1 | 10127.2 | 36145 | 36316 | 1 | 9771.2 | 36866 | 36866 | 1 |
| Portugal | PRT | 2814.132 | 9327 | 9327 | 1 | 2872.246 | 9869 | 9869 | 1 | 2447.702 | 9605 | 9600 | 1 | 2447.702 | 9502 | 9502 | 1 |
| Slovakia | SVK | 2956.663 | 4796 | 4796 | 1 | 3273.747 | 4905 | 4905 | 1 | 3170.945 | 5001 | 5001 | 1 | 3009.8 | 5211 | 5211 | 1 |
| Slovenia | SVN | 6008.357 | 12600 | 12600 | 1 | 6097.567 | 10710 | 10710 | 1 | 5712.463 | 10710 | 10710 | 1 | 5696.8 | 10710 | 10710 | 1 |
| Spain | ESP | 4308.515 | 13046 | 13046 | 1 | 4180.357 | 13046 | 13046 | 1 | 4049.05 | 13046 | 13046 | 1 | 3805.9 | 13046 | 13046 | 1 |
| Switzerland | CHE | 13434.86 | 60413 | 60413 | 1 | 16490.72 | 62555 | 62555 | 1 | 18100.38 | 62820 | 62820 | 1 | 16718.04 | 62845 | 62845 | 1 |
| Czech Repub | CZE | 3621.717 | 147604 | 6002.603 | 24.59 | 3782.041 | 151621 | 6028.908 | 25.149 | 3762.014 | 151338 | 5825.17 | 25.98 | 3762.014 | 154392 | 5606.915 | 27.536 |
| Denmark | DNK | 11871.48 | 199160 | 26730.73 | 7.4506 | 11579.3 | 204880 | 27523.95 | 7.4437 | 11794.5 | 208260 | 27924.75 | 7.4579 | 11620.16 | 211900 | 28424.64 | 7.4548 |
| Hungary | HUN | 2098.396 | 1123200 | 4020.475 | 279.37 | 2091.778 | 1116000 | 3858.254 | 289.25 | 1924.246 | 1176000 | 3961.33 | 296.87 | 1893.005 | 1218000 | 3945.45 | 308.71 |
| Iceland | ISL | 9243.455 | 3055632 | 18929.7 | 161.42 | 9542.277 | 3162576 | 19676.33 | 160.73 | 10422.34 | 3265356 | 20109.35 | 162.38 | 12041.6 | 3382908 | 21844.94 | 154.86 |
| Norway | NOR | 17882.47 | 296585 | 38055.92 | 7.7934 | 19881.49 | 307465 | 41131.89 | 7.4751 | 21277.73 | 319157 | 40882.45 | 7.8067 | 20908.06 | 330857 | 39602.72 | 8.3544 |
| Poland | POL | 1846.015 | 10964 | 2660.778 | 4.1206 | 1959.266 | 11436 | 2732.812 | 4.1847 | 1957.554 | 11860 | 2825.49 | 4.1975 | 2001.138 | 11968 | 2860.216 | 4.1843 |
| Sweden | SWE | 10650.67 | 176800 | 19579.61 | 9.0298 | 11094.92 | 176800 | 20312.27 | 8.7041 | 11893.51 | 176800 | 20435.76 | 8.6515 | 11909.15 | 176800 | 19431.77 | 9.0985 |
| Bulgaria | BGR | 897.6374 | 5084 | 2599.45 | 1.9558 | 828.8408 | 5387 | 2754.372 | 1.9558 | 828.534 | 5686 | 2907.25 | 1.9558 | 882.0043 | 5962 | 3048.369 | 1.9558 |
| Latvia | LVA | 1137.606 | 3631 | 5140.875 | 0.7063 | 1213.032 | 3793 | 5439.553 | 0.6973 | 1377.336 | 3922 | 5590.88 | 0.7015 | 1442.8 | 5898 | 8392.098 | 0.702804 |
| Lithuania | LTU | 1511.149 | 7800 | 2259.036 | 3.4528 | 1599.284 | 7800 | 2259.036 | 3.4528 | 1413.012 | 7800 | 2259.036 | 3.4528 | 1310.578 | 7800 | 2259.036 | 3.4528 |

**Table S4b**

| 2011 | | | | | 2012 | | | | | 2013 | | | | | 2014 | | | | |
|---|---|---|---|---|---|---|---|---|---|---|---|---|---|---|---|---|---|---|---|
| Dependent Variable: μ | | | | | Dependent Variable: μ | | | | | Dependent Variable: μ | | | | | Dependent Variable: μ | | | | |
| Method: Least Squares | | | | | Method: Least Squares | | | | | Method: Least Squares | | | | | Method: Least Squares | | | | |
| Sample: 1 26 | | | | | Sample: 1 26 | | | | | Sample: 1 26 | | | | | Sample: 1 26 | | | | |
| Included observations: 26 | | | | | Included observations: 26 | | | | | Included observations: 26 | | | | | Included observations: 26 | | | | |
| Variable | Coefficient | Std. Error | t-Statistic | Prob. | Variable | Coefficient | Std. Error | t-Statistic | Prob. | Variable | Coefficient | Std. Error | t-Statistic | Prob. | Variable | Coefficient | Std. Error | t-Statistic | Prob. |
| UC | 0.29044 | 0.034564 | 8.40292 | 0 | UC | 0.315257 | 0.030481 | 10.34264 | 0 | UC | 0.330724 | 0.032919 | 10.04648 | 0 | UC | 0.32045 | 0.035267 | 9.086325 | 0 |
| C | 2200.382 | 734.8736 | 2.994233 | 0.0063 | C | 1905.156 | 672.0609 | 2.834797 | 0.0092 | C | 1715.632 | 734.6524 | 2.335297 | 0.0282 | C | 1700.435 | 802.2287 | 2.119638 | 0.0446 |
| R-squared | 0.746325 | | | | R-squared | 0.816752 | | | | R-squared | 0.807895 | | | | R-squared | 0.774778 | | | |
| Adjusted R | 0.735755 | | | | Adjusted R | 0.809117 | | | | Adjusted R | 0.799891 | | | | Adjusted R | 0.765393 | | | |
| Pearson Correlation Coefficient | | t-Statistic | Prob. | | Pearson Correlation Coefficient | | t-Statistic | Prob. | | Pearson Correlation Coefficient | | t-Statistic | Prob. | | Pearson Correlation Coefficient | | t-Statistic | Prob. | |
| 0.863901 | | 8.40292 | 0 | | 0.903743 | | 10.34264 | 0 | | 0.89883 | | 10.04648 | 0 | | 0.880214 | | 9.086325 | 0 | |

**Table S4. Testing results for the data in Fig. 4**. Datasets are composed of 26 European countries in 2011, 2012, 2013 and 2014, and the details of the data are given in Table S4a. $\mu$ (MLCR) and adjusted unemployment compensation (UC) are given in EUR. Annual average exchange rates in the last column of the four separate years are from LCU to EUR. The regression results are shown in Table S4b. The dependent Variable is MLCR, while the independent variable is UC. All parameters are estimated by OLS regression. Standard errors and the $p$-values are reported. Pearson correlation coefficient and the corresponding test statistics, as well as the $p$-values are also reported.

# References

Acemoglu D, Robinson J (2009) Foundation of societal inequality. Science 326 (5953): 678-679

Angle J (1986) The Surplus Theory of Social Stratification and the Size Distribution of Personal Wealth. Social Forces 65: 293-326

Angle J (1992) The Inequality Process and the Distribution of Income to Blacks and Whites. Journal of Mathematical Sociology 17: 77-98

Angle J (1993) Deriving the Size Distribution of Personal Wealth from "The Rich Get Richer, the Poor Get Poorer", Journal of Mathematical Sociology 18: 27-46

Angle J (1996) How the Gamma Law of Income Distribution Appears Invariant under Aggregation. Journal of Mathematical Sociology 31: 325-358

Angle J (2006) The Inequality Process as a wealth maximizing process. Physica A 367: 388-414

Arrow K J (1963) Social choice and individual values (John Wiley& Sons, Inc., New York)

Arrow K J, Debreu G (1954) Existence of an equilibrium for a competitive economy. Econometrica 22 (3): 265–290

Atkinson A B, Piketty T, Saez E (2011) Top incomes in the long run of history. Journal of Economic Literature 49 (1): 3–71

Autor D H (2014) Skills, education, and the rise of earnings inequality among the "other 99 percent". Science 344 (6186): 843–851

Autor D H, Katz L F, Kearney M S (2008) Trends in U.S. wage inequality: revising the revisionists. Review of Economics and Statistics 90 (2): 300–323

Axtell R L (2001) Zipf distribution of U.S. firm sizes. Science 293 (5536) 1818–1820

Banerjee A, Yakovenko V M (2010) Universal patterns of inequality. New J. Phys. 12: 075032

Banerjee A, Yakovenko V M, Di Matteo T (2006) A study of the personal income distribution in Australia. Physica A 370 (1) 54–59

Chakrabarti A S, Chakrabarti B K (2009) Microeconomics of the ideal gas like market models. Physica A 388 (19): 4151–4158

Chakrabarti B K, Chakraborti A, Chakravarty S R, Chatterjee A (2013) Econophysics of income and wealth distributions (Cambridge University Press)

Cho A (2014) Physicists say it's simple. Science 344 (6186): 828–828

Clementi F, Gallegati M, Kaniadakis G (2010) A model of personal income distribution with application to Italian data. Empirical Economics 39: 559-591

Clementi F, Gallegati M, Kaniadakis G (2012) A new model of income distribution: the κ-generalized distribution. Journal of Economics 105: 63-91

Derzsy N, Néda Z, Santos M A (2012) Income distribution patterns from a complete social security database. Physica A 391 (22): 5611–5619

Dopfer K (2004) The economic agent as rule maker and rule user: Homo Sapiens Oeconomicus. Journal of Evolutionary Economics 14: 177–195

Dragulescu A, Yakovenko V M (2000) Statistical mechanics of money. Eur. Phys. J. B 17 (4): 723–729

Dragulescu A, Yakovenko V M (2001a) Evidence for the exponential distribution of income in the USA. Eur. Phys. J. B 20 (4): 585–589

Dragulescu A, Yakovenko V M (2001b) Exponential and power-law probability distributions of wealth and income in the United Kingdom and the United States. Physica A 299 (1-2): 213–221

Foley D K (1994) A statistical equilibrium theory of markets. Journal of Economic Theory 62 (2): 321–345

Foster J, Metcalfe J S (2012) Economic emergence: An evolutionary economic perspective. Journal of Economic Behavior & Organization 82 (2-3): 420–432

Golosov M, Maziero P, Menzio G (2013) Taxation and redistribution of residual income inequality. Journal of Political Economy 121(6): 1160–1204

Harte J, Zillio T, Conlisk E, Smith A B (2008) Maximum entropy and the state-variable approach to macroecology. Ecology, 89 (10): 2700–2711

Heathcote J, Storesletten K, Violante G L (2010) The Macroeconomic Implications of Rising Wage Inequality in the United States. Journal of Political Economy 118(4): 681–722

Hodgson G M (2004) The Evolution of Institutional Economics: Agency, Structure and Darwinism in American Institutionalism. (Routledge, London and New York)

Jagielski M, Kutner R (2013) Modelling of income distribution in the European Union with the Kokker-Planck equation. Physica A 392 (9): 2130–2138

Jones C I (2015) Pareto and Piketty: The Macroeconomics of Top Income and Wealth Inequality. Journal of Economic Perspectives 29(1): 29–46

Kakwani N (1980) Income inequality and poverty (Oxford University Press)

Katz L, Autor D (1999) Changes in the Wage Structure and Earnings Inequality, in Handbook of Labor Economics, O. Ashenfelter and D. Card, eds. (Amsterdam: North-Holland), Vol. 3A.

Kuznets S (1955) Economic growth and income inequality. American Economic Review 45 (1): 1–28

Lai T L, Robbins H, Wei C Z (1979) Strong consistency of least squares estimates in multiple regression II. Journal of Multivariate Analysis 9 (3): 343–361

Lambert P J (1993) The distribution and redistribution of income: A mathematical analysis (2ed). (Manchester University Press)

Lux T, Marchesi M (1999) Scaling and criticality in a stochastic multi-agent model of a financial market, Nature 397: 498-500

Mackmurdo A H (1940) The social organism. Nature 145 (3666): 187–187

Mandelbrot B (1960) The Pareto-Levy law and the distribution of income. International Economic Review 1 (2): 79-106

Mas-Collel A, Whinston M D, Green J R (1995) Microeconomic theory (Oxford University Press)

Moretti E (2013) Real Wage Inequality. American Economic Journal: Applied Economics 5(1): 65–103

Nelson R R, Winter S G (1982) An evolutionary theory of economic change (The Belknap Press of Harvard University Press)

Nirei M, Souma W (2007) A two factor model of income distribution dynamics. Review of

Income and Wealth 53 (3): 440–459

Nishi A, Shirado H, Rand D G, Christakis N A (2015) Inequality and visibility of wealth in experimental social networks. Nature 526 (7573): 426–429

Oancea B, Andrei T, Pirjol D (2016) Income inequality in Romania: the exponential-Pareto distribution. Physica A, DOI: http://dx.doi.org/10.1016/j.physa.2016.11.094

Pareto V (1897) Cours d' Economie Politique (L' Universite de Lausanne)

Piketty T (2003) Income Inequality in France, 1901-1998. Journal of Political Economy 111: 1004–1042

Piketty T, Qian N (2009) Income Inequality and Progressive Income Taxation in China and India, 1986–2015. American Economic Journal: Applied Economics 1 (2): 53–63

Piketty T, Saez E (2003) Income inequality in the United States, 1913 - 1998. Quarterly Journal of Economics 118: 1–39.

Piketty T, Saez E (2014) Inequality in the long run. Science 344 (6186): 838–843

Potts J (2001) Knowledge and markets. Journal of Evolutionary Economics 11: 413–431

Ravallion M (2014) Income inequality in the developing world. Science 344 (6186): 851–855

Rawls J (1999) A theory of justice (Revised Edition) (Harvard University Press)

Rudin W (1976) Principles of Mathematical Analysis (Third Edition) (McGraw-Hill, Inc.)

Saez E, Zucman G (2016) Wealth Inequality in the United States since 1913: Evidence from Capitalized Income Tax Data. Quarterly Journal of Economics 131: 519–578

Shaikh A (2016) Income distribution, econophysics and Piketty. Rev. Polit. Econ., DOI:10.1080/09538259.2016.1205295

Shaikh A, Papanikolaou N, Wiener N (2014) Race, gender and the econophysics of income distribution in the USA. Physica A 415: 54–60

Silva A C, Yakovenko V M (2005) Temporal evolution of the "thermal" and "superthermal" income classes in the USA during 1983-2001. Europhys. Lett. 69 (2): 304–310

Tao Y (2010) Competitive market for multiple firms and economic crisis. Phys. Rev. E 82 (3): 036118

Tao Y (2015) Universal laws of human society's income distribution. Physica A 435: 89–94

Tao Y (2016) Spontaneous economic order, Journal of Evolutionary Economics 26 (3): 467–500

Tao Y (2017) An Index Measuring the Deviation of a Real Economy from the General Equilibrium: Evidence from the OECD Countries, Available at SSRN: http://dx.doi.org/10.2139/ssrn.2792556

Tao Y, Wu X, Li C (2017) Rawls' fairness, income distribution and alarming level of Gini coefficient. Economics Discussion Papers, No 2017-67, Kiel Institute for the World Economy. http://www.economics-ejournal.org/economics/discussionpapers/2017-67

Venkatasubramanian V, Luo Y, Sethuraman J (2015) How much inequality in income is fair? A microeconomic game theoretic perspective. Physica A 435: 120–138

Walras L (2003) Elements of pure economics or the theory of social wealth (Routledge)

Whitfield J (2007) Survival of the likeliest? PLoS Biol 5 (5): e142

Yakovenko V M, Rosser Jr. J B (2009) Statistical mechanics of money, wealth, and income. Review of Modern Physics 81 (4): 1703–1717